\documentclass[aps,prx,twocolumn,superscriptaddress]{revtex4-2}
\usepackage{amsmath,amsfonts}
\usepackage[dvipsnames]{xcolor}
\usepackage{graphicx}
\usepackage[braket, qm]{qcircuit}
\usepackage{dcolumn}
\usepackage{bm}
\usepackage{braket}

\newcommand{\gates}[1]{\ensuremath{\mathsf {#1}}}

\usepackage{url}

\newcommand{\tr}{\operatorname{Tr}}

\newcommand{\Yorktown}{IBM Quantum, IBM T.J. Watson Research Center, Yorktown Heights, NY 10598, USA}
\newcommand{\Almadan}{IBM Quantum, IBM Almaden Research Center, San Jose, CA 95120, USA}

\begin{document}

\title{Error mitigation with stabilized noise in superconducting quantum processors}
\author{Youngseok Kim}
\affiliation{\Yorktown}
\author{Luke C. G. Govia}
\affiliation{\Almadan}
\author{Andrew Dane}
\affiliation{\Yorktown}
\author{Ewout van den Berg}
\affiliation{\Yorktown}
\author{David M. Zajac}
\affiliation{\Yorktown}
\author{Bradley Mitchell}
\affiliation{\Almadan}
\author{Yinyu Liu}
\affiliation{\Yorktown}
\author{Karthik Balakrishnan}
\affiliation{\Yorktown}
\author{George Keefe}
\affiliation{\Yorktown}
\author{Adam Stabile}
\affiliation{\Yorktown}
\author{Emily Pritchett}
\affiliation{\Yorktown}
\author{Jiri Stehlik}
\affiliation{\Yorktown}
\author{Abhinav Kandala}
\affiliation{\Yorktown}
\date{\today}

\begin{abstract}

    Pre-fault tolerant quantum computers have already demonstrated the ability to estimate observable values accurately, at a scale beyond brute-force classical computation. This has been enabled by error mitigation techniques that often rely on a representative model of the device noise. However, learning and maintaining these models is complicated by fluctuations in the noise over unpredictable time scales, for instance, arising from resonant interactions between superconducting qubits and defect two-level systems (TLS). Such interactions affect the stability and uniformity of device performance as a whole, but also affect the noise model accuracy, leading to incorrect observable estimation. Here, we experimentally demonstrate that tuning of the qubit-TLS interactions helps reduce noise instabilities and consequently enables more reliable error-mitigation performance. These experiments provide a controlled platform for studying the performance of error mitigation in the presence of quasi-static noise. We anticipate that the capabilities introduced here will be crucial for the exploration of quantum applications on solid-state processors at non-trivial scales.

\end{abstract}

\maketitle

\section{Introduction}

A common primitive in many near-term quantum algorithms is the accurate estimation of observable expectation values for a short-depth quantum circuit~\cite{preskill2018quantum,bharti2022noisy}. Since, in practice, these circuits are run on noisy quantum processors, the observable estimates tend to be biased away from the true values. In the absence of fault tolerance, quantum error mitigation methods provide a viable way to access more accurate observable estimates~\cite{temme2017error,Li2017Efficient,kandala2019error,kim2023scalable,kim2023evidence,berg2023probabilistic,filippov2023scalable}. These methods typically rely on combining the results of several noisy quantum circuits in ways that cancel the effect of noise on observable estimates, with no qubit overhead. One of these methods, namely zero-noise extrapolation (ZNE), was recently demonstrated to work on circuit sizes that are beyond brute-force classical simulation~\cite{kim2023evidence}.

Several error-mitigation methods rely on access to a representative model of the device noise~\cite{temme2017error,berg2023probabilistic,kim2023evidence}. Learning and maintaining an accurate model, however, is complicated by fluctuations in physical device noise, which can occur over unpredictable time scales. In superconducting quantum processors, one prominent source of such fluctuations is the interaction between qubits and defect two-level systems (TLS)~\cite{burnett2014evidence,lisenfeld2015observation,Graaf2017direct,Klimov2018fluctuations,de2020two,carroll2022dynamics,thorbeck2023tls}. It has been shown that the diffusion of TLS transition frequencies over time is a major contributor to fluctuations in qubit relaxation times~\cite{Klimov2018fluctuations,carroll2022dynamics,thorbeck2024prl}, and these dynamics have been extensively investigated through a number of experimental controls that modulate the qubit-TLS interaction, including: external electric fields~\cite{Sarabi2016,lisenfeld2019electric}, structural deformation of the physical processor~\cite{grabovskij2012strain,lisenfeld2015observation}, flux tuning~\cite{Klimov2018fluctuations}, and microwave drives~\cite{carroll2022dynamics,Abdurakhimov2020,thorbeck2024prl}.  Overall, fluctuations in qubit relaxation times have crucial consequences for the stability~\cite{Martinez21,dasgupta2023,dasgupta2024}, uniformity, and throughput of superconducting quantum computation. As an example, previous studies have shown that qubit-TLS interaction can degrade the performance of error-mitigation ~\cite{kim2023evidence}, or even result in unphysical observable estimates~\cite{kim2023scalable}.

In this work, we present experiments that focus on noise instabilities in superconducting qubit hardware and their impact on the performance of error mitigation. We employ a device with six \textcolor{black}{fixed-frequency} transmon qubits connected in a one-dimensional chain \textcolor{black}{(Q1-Q2-$\cdots$-Q6)} with tunable couplers~\cite{McKay2016,Yan2018,Mundada2019,Foxen2020,Stehlik2021}.
\textcolor{black}{Each transmon qubit is equipped with an electrode that is placed above to modulate the qubit-TLS interaction. Applying a bias on the electrodes, parametrized by $k_{TLS}$ in arbitrary units, modifies the local electric field at defect sites and modulates the TLS resonance frequency~\cite{lisenfeld2019electric}.  Therefore, representative characteristics such as $T_1$ can be modulated by $k_{TLS}$ due to subsequent changes in qubit-TLS interaction.} We demonstrate that, on average, controlling the qubit-TLS interaction can help improve and even stabilize worst-case $T_1$ instances. We then study the impact these results have on noise characteristics and gate layer performance. Finally, we leverage improved stability of qubit noise in the context of observable estimation using quantum error mitigation.

\begin{figure*}[!t]
\includegraphics[width=1.5\columnwidth]{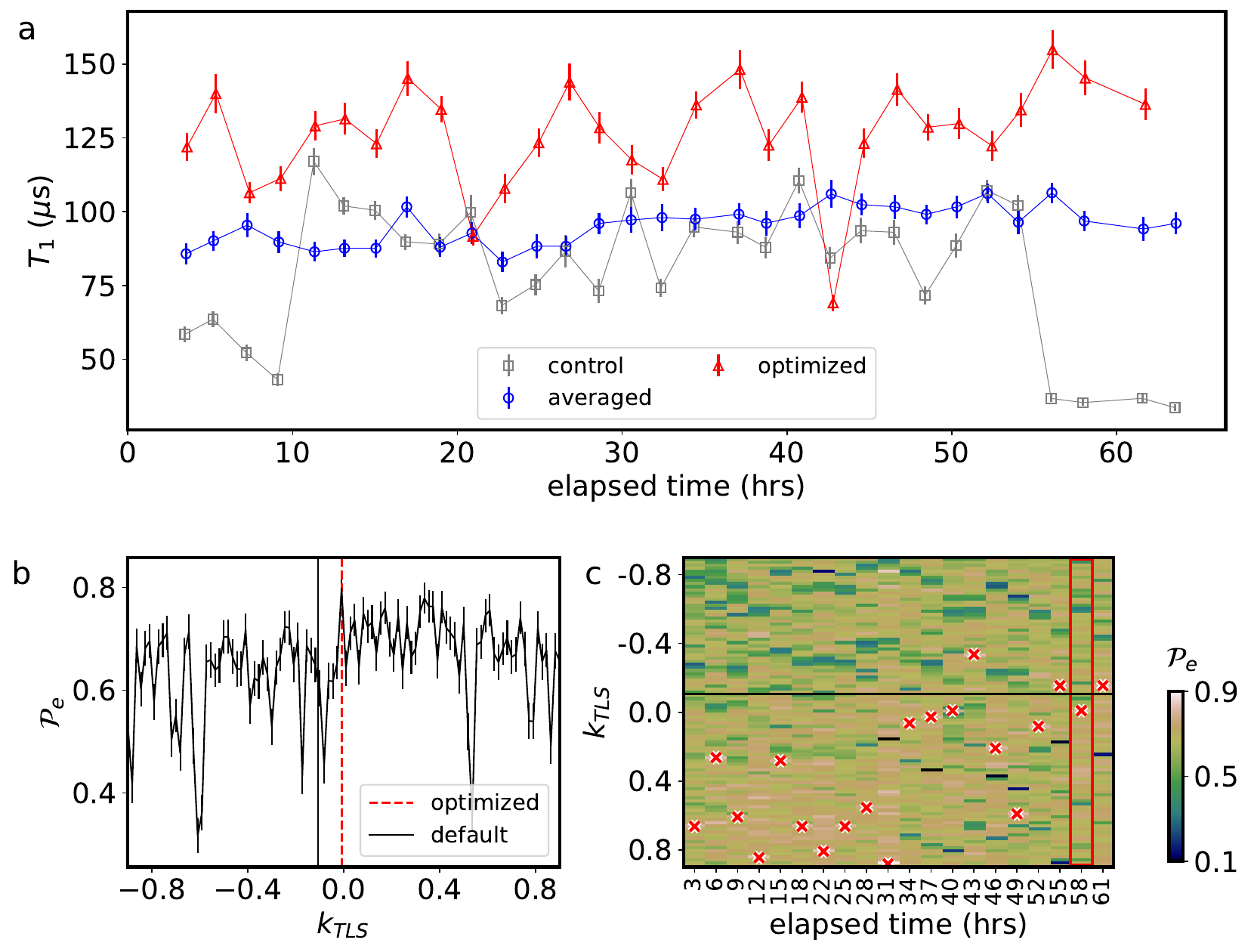}

\caption{{\bf Stabilizing $T_1$} (a) Experimentally measured $T_1$ for a given $k_{TLS}$.  Black rectangles show $T_1$ fluctuation without any optimization procedure. Red triangles illustrate $T_1$ after optimization is carried out as illustrated in (b). Blue circles show $T_1$ measured by averaging over the TLS landscapes. Averaging is achieved by modulating $k_{TLS}$ with 1Hz sine wave with amplitude 0.5 on top of the default value. The corresponding $T_1$ decay curves are shown in the Supplementary material. (b) At each $k_{TLS}$, we prepare an excited state for a given qubit and then probe the probability of measuring the excited state, $\mathcal{P}_e$, after $40\mu$s. Ideally, the qubit would stay in the excited state, however, $T_1$ decay results in $\mathcal{P}_e < 1$~\cite{carroll2022dynamics}.  The decay of $\mathcal{P}_e$ therefore serves as a proxy for $T_1$ decay. By changing the magnitude $k_{TLS}$ of the TLS modulation we can alter the qubit-TLS interaction strength. This causes $\mathcal{P}_e$ to vary as a function of $k_{TLS}$. When a strong qubit-TLS interaction exists, a pronounced decay of $\mathcal{P}_e$ is observed~\cite{carroll2022dynamics}, as seen by the dips in the curve.  The vertical lines indicate the choice of $k_{TLS}$ that maximizes $\mathcal{P}_e$ (dashed), compared to an arbitrarily chosen default $k_{TLS}$ value (solid). (c) We monitor the TLS landscape over time to illustrate the fluctuation of qubit-TLS interaction dynamics. The black horizontal line indicates $k_{TLS}$ of the control experiment, the red crosses indicate the values used for the optimized experiment. The red rectangular box indicates the data illustrated in (b). The experiments are carried out with 1kHz repetition rate and values are obtained from 300 single shot measurements.}\label{fig:t1}
\end{figure*}

\section{Stabilizing $T_1$}

In this section, we focus on temporal fluctuations of the qubit-TLS interaction and different strategies to minimize its impact on $T_1$. Over a $60$ hours period, the qubit $T_1$ values on the device are seen to fluctuate on average by over 300\% (see Supplementary~\ref{app:device}). Figure~\ref{fig:t1}a depicts an example of these fluctuations for one of these qubits.
We now consider modulation of the qubit-TLS interaction by means of a control parameter $k_{TLS}$. To study its effect, we use the excited state population of the qubit, $\mathcal{P}_e$, measured after a fixed delay time of $40\mu$s as a quick proxy for $T_1$~\cite{carroll2022dynamics}. The resulting $\mathcal{P}_e$ values obtained for a range of $k_{TLS}$ parameters are shown in Fig.~\ref{fig:t1}b. The peaks and dips in the plot are a reflection of the qubit-TLS interaction landscape as modulated by $k_{TLS}$ at the given time instance. \textcolor{black}{Meantime, we observe that no further calibration is needed for qubit itself as we mainly modulate TLS characteristics unless we had a pronounced qubit-TLS interaction during initial qubit calibration.}
Repeating the same experiment at different times, Fig.~\ref{fig:t1}c illustrates the temporal fluctuation of the qubit-TLS interaction.

The value of the $k_{TLS}$ parameter clearly has a strong effect on $\mathcal{P}_e$, which raises the natural question of how best to select it. One way to improve qubit coherence is to actively monitor the temporal snapshot of the TLS landscape and choose $k_{TLS}$ that produces the best $\mathcal{P}_e$. The results show a clear benefit from this optimization, improving the overall $T_1$ in Fig.~\ref{fig:t1}a. This strategy, which we refer to as the \emph{optimized} noise strategy, requires active monitoring of the TLS environment. Between monitoring events, the qubit remains exposed to random fluctuations in the qubit-TLS interaction. Alternatively, we mitigate the impact of these fluctuations by averaging over randomly sampled TLS environments per shot. This is achieved by applying slowly varying sinusoidal (or triangular) amplitude modulation on $k_{TLS}$. The applied modulation frequency (1Hz) is much lower than the shot repetition rate (1kHz). Therefore, the modulation is effectively quasi-static within each shot, but samples a different quasi-static TLS environments for each shot over the duration of entire experiment. We refer to this approach as the \emph{averaged} noise strategy. Figure~\ref{fig:t1}a illustrates that the averaged $T_1$ value is more stable than that of the control and optimized experiments. Since this strategy only requires passive sampling of the TLS environment from shot-to-shot, it does not require constant monitoring. A natural question to consider is whether the $T_1$ decay is well approximated by a single-exponential in this case -- this is seen to hold for this data set (see Supplementary Fig.~\ref{fig:t1supp}). However, this may not always be the case, as is often seen in scenarios with strong couplings to TLS, and the implications of this for noise learning and mitigation are discussed later and detailed in Supplementary~\ref{app:theory}.

\begin{figure*}[!ht]
    \includegraphics[width=2\columnwidth]{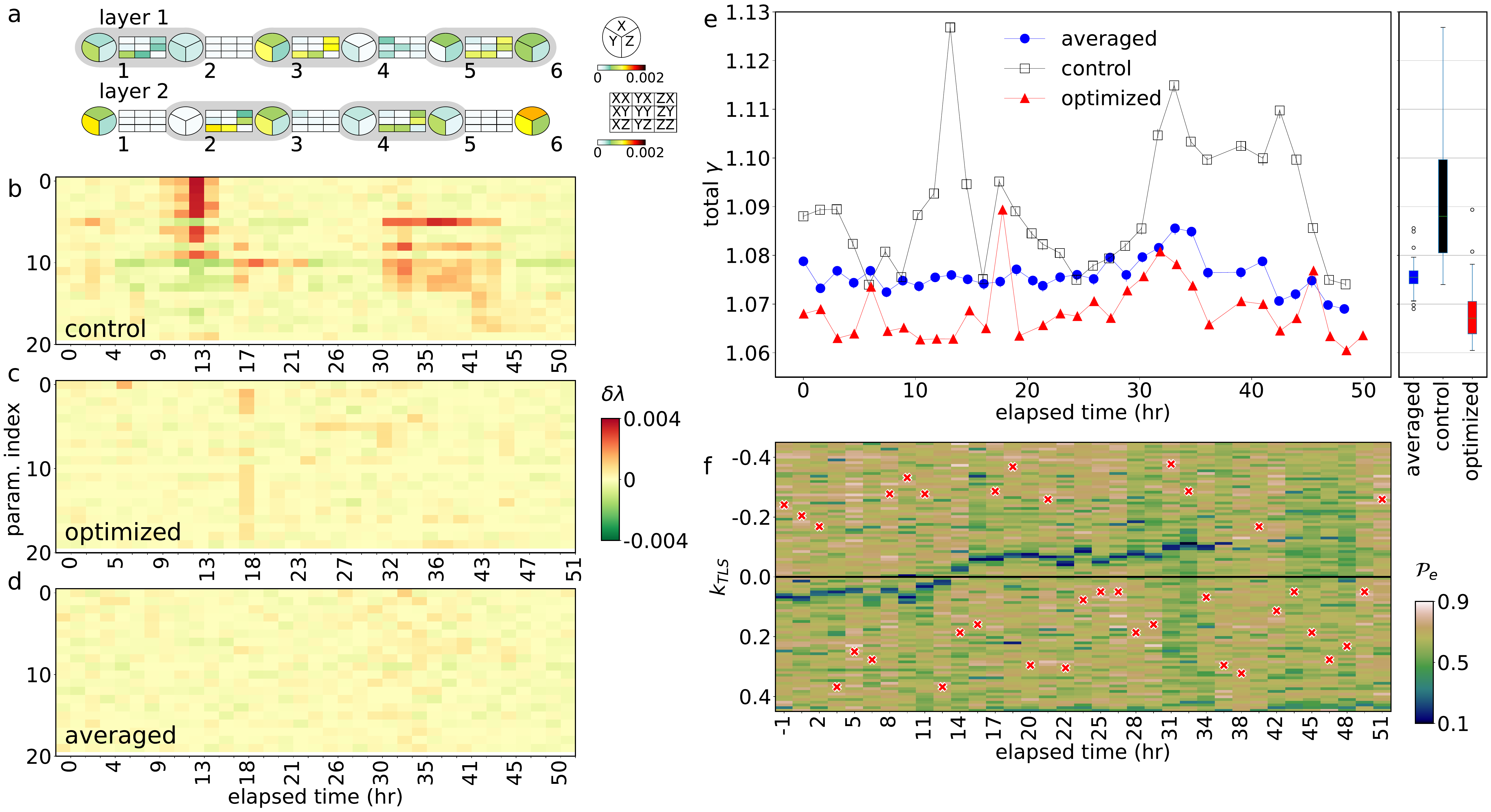}

\caption{{\bf Stabilizing noise in gate layers} (a) Graphical representation of the coefficients of the sparse Pauli-Lindblad noise models of two layers. The layers consist of $\gates{CZ}$ gates covering different qubit pairs (shaded box): $\{(1,2),(3,4),(5,6)\}$ for layer 1 and $\{(2,3),(4,5)\}$ for layer 2. The model parameters apply to Pauli terms on each of the six qubits, as well as weight-two Pauli terms on connected qubits. The model parameters $\{\lambda_k\}_{k\in{\mathcal{K}}}$ are determined by applying the learning protocol separately to each of the two layers. The inset on the right depicts the position of the Pauli coefficients and the color bars for the one- and two-qubit terms. (b--d) Provides a more detailed picture of model parameter instability by computing the model coefficient fluctuation $\delta\lambda_k(t)=\lambda_k(t)-\text{median}[\lambda_k(t)]$, where $\text{median}[\cdot]$ computes a median value of time varying model coefficient. The plot shows $\delta\lambda_k(t)$ at specific time $t$ (x-axis) for various Pauli jump terms $P_k$ (y-axis). We show the first 20 parameters sorted by maximum fluctuation. (e) Shows the total sampling overhead, $\gamma$, monitored over time for three different scenarios: (i) a baseline \emph{control} experiment with $k_{TLS}=\kappa$ held at a constant neutral point $\kappa$ for all qubits; (ii) an \emph{averaged} experiment where $k_{TLS}$ is set to $\kappa$ plus a slowly varying triangular wave with 1Hz frequency and amplitude of $\pm 0.2$; and (iii) an \emph{optimized} experiment carried out by periodically update $k_{TLS}$ to a value that maximizes $\mathcal{P}_e$. Optimization is performed just prior to each learning experiment for the optimized noise channel. The right inset illustrates the median value, and first and third quartile of each experiment.  (f) For the optimized experiment, the qubit-TLS interaction landscape of Q2 is probed using TLS control parameter $k_{TLS}$. The plot displays the resulting $\mathcal{P}_e$ over time, along with the optimized control parameters, indicated by the red crosses.  Strong qubit-TLS interactions appear as dark green boxes, and can be seen to drift close to the neutral point $\kappa$ (horizontal black line) at an elapsed time of $\sim13$ hr. This coincides with the elevated noise level $\gamma$ and noticeable larger fluctuations over time in plot (e). The data used for this plot is the same as that used for plot (b--d). Error bars in (e) are determined using 100 bootstrapped instances of the experimental data. Likewise, error bars for $\lambda_k$ are obtained from 100 bootstrapped instances and maximum fluctuation of each row in (b--d) are all larger than the error bar. Further details on experiment conditions are described in Supplementary~\ref{app:getlambda}.}
    \label{fig:model}
\end{figure*}

\section{Stabilizing noise in gate layers}

Having demonstrated different modulation strategies to stabilize $T_1$, we now extend our study to the characterization of noise associated with layers of concurrent entangling two-qubit gates. An accurate characterization of the noise enables us to remove its effect on observable estimation by means of probabilistic error cancellation (PEC)~\cite{temme2017error}. Previous experiments have shown that noise on our devices can be tailored in such a way that a sparse Pauli-Lindblad (SPL) model~\cite{berg2023probabilistic} accurately captures the noise and facilitates the estimation of accurate observable values using PEC, and even enables ZNE experiments exceeding 100 qubits~\cite{kim2023evidence}. This motivates us to study the impact on the optimized and averaged noise strategies have on the learned SPL model parameters and their stability.

The sparse Pauli-Lindblad noise model proposed in~\cite{berg2023probabilistic} provides a scalable framework for learning the noise associated with a layer of gates. The sparsity of the noise model is achieved by imposing assumptions on the noise in real hardware. \textcolor{black}{We tailor and learn the noise employing protocols described in ref~\cite{berg2023probabilistic} and references therein.} First, by applying Pauli twirling, we ensure that the noise can be described by a Pauli channel. We then model the noise as $\mathcal{E}(\rho) =\exp[\mathcal{L}](\rho)$, where $\mathcal{L}$ represents a Lindbladian with Pauli jump terms $P_k$ weighted by non-negative model coefficients $\lambda_k$. Second, we can obtain a sparse noise model by making the reasonable assumption that noise originates locally on individual or connected pairs of qubits. This allows us to restrict the set of generators $\mathcal{K}$ to one- and two-local Pauli terms in accordance with the qubit topology. The model parameters $\lambda_k$ are characterized by measuring the channel fidelities of Pauli operators using a procedure described in Supplementary~\ref{app:getlambda}. 
The fact that the individual $P_k$ terms in $\mathcal{L}$ commute make the noise model ideal for probabilistic error cancellation~\cite{temme2017error,berg2023probabilistic}, since the channels generated by each of these terms can be inverted independently. The inverse channels, however, are non-physical and the observable is therefore reconstructed through post-processing. In the absence of model inaccuracy, we obtain an unbiased estimate for the expectation value of any Pauli operator. The variance of the estimator, however, is amplified by a multiplicative factor $\gamma=\exp\Big(\sum_{k\in\mathcal{K}} 2\lambda_k\Big)$, which can be compensated by increasing the number of samples~\cite{berg2023probabilistic}. We therefore refer to $\gamma$ as the \emph{sampling overhead}. In the following, we experimentally monitor individual model parameters, $\lambda_k$, and connect the overall noise strength to runtime overhead for error mitigation by using the sampling overhead, $\gamma$.

We now experimentally characterize the optimized and averaged noise channels over time with a goal of assessing whether the TLS modulation strategies succeed in stabilizing the noise, and therefore the model parameters. For this, we learn model parameters $\lambda_k$ associated with two different gate layers for a one-dimensional chain of six qubits, covering all local \gates{CZ} gate pairs. Figure~\ref{fig:model}a shows the parameter values obtained for the optimized noise channel for a single learning experiment.  To quantify fluctuations in the noise, we repeat the noise characterization over time and track the learned model parameters over $\sim$50 hours of monitoring for the control, optimized, and averaged noise channels in Fig.~\ref{fig:model}b-d.

The control experiment depicts large fluctuations in the model parameters around 13 hours of elapsed time, which shows reasonable correlation with a strong Q2-TLS interaction occurring around the same time (Figure~\ref{fig:model}f). 
Meanwhile, the optimized experiment selects $k_{TLS}$ that aims to avoid impact of nearby qubit-TLS impact while maximizing $\mathcal{P}_e$, indicated by red cross symbols in Fig.~\ref{fig:model}f. This optimization is performed right before learning experiment and help us to avoid configurations that have particularly strong qubit-TLS interaction. Aside from the relatively smaller aberrations associated with short term fluctuations that are fixed by the next optimization round, the model parameters are seen to be largely stable over the duration of the experiment, Fig.~\ref{fig:model}c. The smaller aberrations are seen to be further stabilized in the averaged case, Fig.~\ref{fig:model}d.

The time dependence of the model parameters can be further extended to track the stability of the total sampling overhead $\gamma$, which is given by the product of $\gamma_1$ and $\gamma_2$, the sampling overhead values for each of the two layers, respectively. Figure~\ref{fig:model}e shows the optimized strategy attains the lowest overall sampling overhead. Finally, the averaged noise channel experiment exhibits better stability compared to the control experiment. In addition, the stability we obtain from this scheme does not require frequent monitoring, and we expect more stable operation during 
periods when the experiment is running and monitoring jobs are not.

Additional analysis in Fig.~\ref{fig:model}b--f shows that the observed stability $\gamma$ reflects the stability of single- and two-qubit model parameters. This is an important observation since temporal fluctuations or noise control strategies may induce discrepancies between learned parameters at one time and actual parameters at a later time, which would greatly reduce the ability to perform meaningful error mitigation. For instance, one might worry that, after each $T_1$ optimization stage, the new $T_1$ and, along with it, the noise channel may be completely different. Figure~\ref{fig:model}c experimentally shows that this is largely not the case. 
Aside from the relatively smaller aberrations associated with short-term fluctuations that are fixed by the next optimization round, most of the model coefficients remain relatively consistent for most of the time. Figure~\ref{fig:model}d similarly visualizes that the averaged noise models exhibit consistent model parameter values throughout the duration of the experiment.

\begin{figure}[!t]
    \includegraphics[width=0.98\columnwidth]{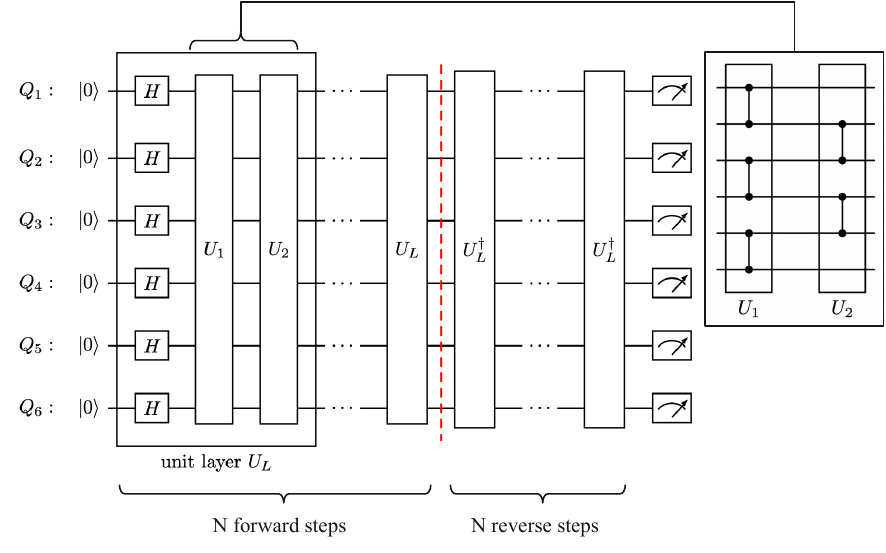}

\caption{{\bf Benchmark circuit for error mitigation.}  We use a 6-qubit mirror circuit to benchmark the error-mitigation performance. The circuit features two unique layers, $U_1$ and $U_2$, of two-qubit entangling $\gates{RZZ}(\pi/2)=\gates{CZ}$ gates, as shown in the inset on the right. We next define a compound block $U_L$ of single-qubit Hadamard gates on each qubit, followed by the two layers. The benchmark circuit is then constructed by repeating the $U_L$ circuit $N=10$ times, followed by an equal number of reverse operations $U_L^{\dag}$.
}
    \label{fig:circuit}
\end{figure}

\begin{figure*}[t]
    \includegraphics[width=2.0\columnwidth]{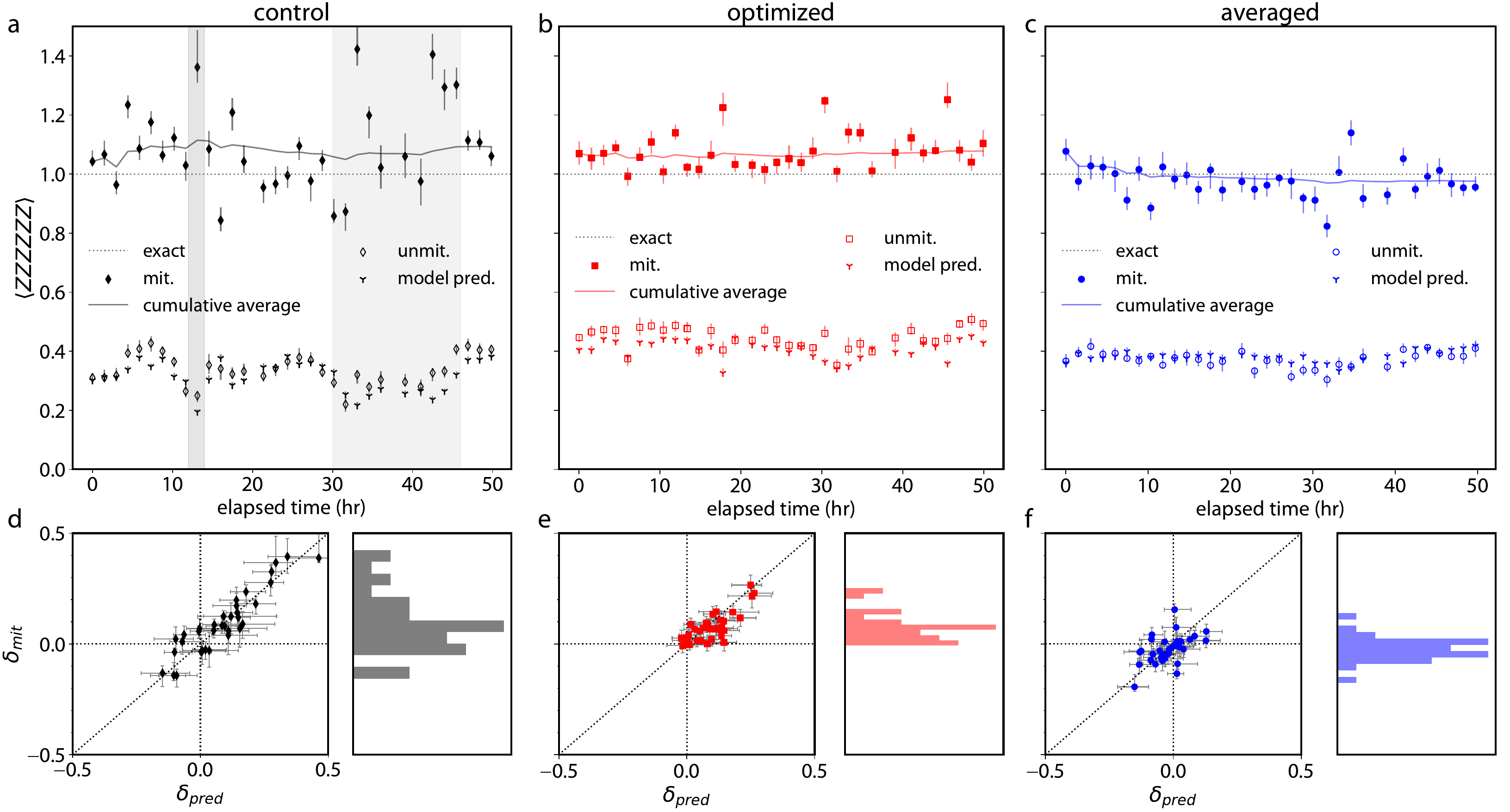}

    \caption{{\bf Error mitigation results for different qubit-TLS modulation strategies.} (a-c) Weight-6 observable ($\braket{ZZZZZZ}$) estimates as a function of time using the three different strategies with (filled markers) and without (open markers) error mitigation, along with the cumulative average of the mitigated observable values (solid line). The experiment is performed following the schedule described in Fig.~\ref{fig:model} for the (a) control, (b) optimized, and (c) averaged modulation strategies. For reference, the ideal observable of $1$ is indicated as a dashed line. The mitigated observables can be seen to fluctuate near the ideal value, and the shaded regions highlight time windows with high fluctuations as determined by the analysis in Fig.~\ref{fig:model}d. Each data point is obtained from 4096 random circuit instances with 32 shots per circuit. During the experiment, we interleave 2048 random circuit instances for readout-error mitigation~\cite{berg2022model} and 512 random circuit instances for estimating the unmitigated observable. Readout-error mitigation is applied to both the unmitigated and mitigated observable estimates. Error bars for the unmitigated and mitigated results are obtained by bootstrapping the PEC result 25 times. (d-f) Scatter plots of the predicted ($\delta_{pred}$) and observed ($\delta_{mit}$) deviations of the observable from the ideal expectation value. The correlation between $\delta_{pred}$ and $\delta_{mit}$ confirms that the temporal fluctuation of the noise model plays a role in mitigation error observed in PEC protocol. The histograms along the y-axes show the respective distributions of $\delta_{mit}$.}
\label{fig:mit}
\end{figure*}

\section{Stability of quantum error mitigation}

In the previous section it was shown that noise models for the optimized and averaged noise channels remain consistent over time.  We now study their impact on error mitigation of a benchmark circuit. The circuit we use is a mirror circuit on a chain of six qubits with alternating layers of $\gates{CZ}$ gates that cover all neighboring qubit pairs, as illustrated in Fig.~\ref{fig:circuit}. Since the circuit is mirrored it effectively implements an identity operator, the ideal expectation value of all Pauli-Z observables is equal to 1. Each experiment consists of a noise learning and a mitigation stage. During the first stage we learn the noise associated with the two layers of $\gates{CZ}$ gates for each of the control, optimized, and averaged noise channel setting. We then run error mitigation using the learned noise models in their respective setting. For each stage we interleave the circuits for all three settings to ensure they all run within the same time window.

Figures~\ref{fig:mit}a--c show the mitigated and unmitigated values for the weight-6 observable $\braket{ZZZZZZ}$ for the control, optimized, and averaged noise channels, respectively, with independent runs over a $\sim 50$ hour period.  In all three setting, the unmitigated values are $0.341\pm0.052$, $0.446\pm0.036$, and $0.371\pm0.027$ for control, optimized, and averaged, respectively -- a large deviation from the ideal value of 1. While fluctuations in the mitigated observable values are seen in all three setting, they are most pronounced in the control setting, where they correlate with periods of strong TLS interaction, as shown earlier in Fig.~\ref{fig:model}f. In principle, one could run separate background circuits to monitor TLS interactions and trigger disposal of data acquired during periods of large fluctuation to improve performance. However, the limited shot budget for such monitoring circuits greatly reduces the ability to accurately detect all but the largest fluctuation events, as discussed in Supplementary ~\ref{app:expstrategy} and~\ref{app:controlonly}. Another strategy to reduce fluctuations in the observable values is to average the results of multiple independent learning-mitigation cycles, as seen from the cumulative average trace of Fig.~\ref{fig:mit}a. However, such an approach is increasingly expensive for larger circuits, as discussed in Supplementary~\ref{app:expstrategy}. Meanwhile, Figs.~\ref{fig:mit}b and~\ref{fig:mit}c demonstrate that both the optimized and averaged noise strategies help stabilize the error mitigation results and enable smaller fluctuations than observed in the control experiment. The improvement is predicted to be even more prominent at larger depths, see Supplementary Fig.~\ref{fig:histdeeper}.

A leading source of fluctuations in the mitigated results in all three settings could be attributed to drifts in the device noise between the noise-learning step and the subsequent mitigation step.
To quantify such deviations, without relearning the noise model, we proceed as follows. First, note that Clifford circuits subject to Pauli noise can be simulated efficiently. This allows us to predict the expected noisy observable value $\langle \tilde{O}\rangle_{pred} = f_{pred}\langle O\rangle$.  Second, by running unmitigated benchmark circuits interleaved with the mitigation circuits, we can measure both the noisy observable value $\langle \tilde{O}\rangle = f_{exp}\langle O\rangle$, and the error-mitigated estimate $\langle O\rangle_{mit}$.  In the absence of noise-model fluctuations, we have $f_{exp}=f_{pred}$, which allows us to recover the ideal observable $\braket{O}=\braket{\tilde{O}}/f_{pred}$. In the presence of the noise fluctuations, however, this no longer holds \textcolor{black}{and may lead to under- or over-estimation on the target observable}. We quantify this known source of deviation on the mitigated observable as $\delta_{pred}=\braket{\tilde{O}}/f_{pred}-1$. In Fig.~\ref{fig:mit}d, we plot an expected deviation due to the noise fluctuation ($\delta_{pred}$) in x-axis and an observed deviation ($\delta_{mit}=\braket{O}_{mit}-1$) in y-axis. The plot shows a clear correlation between $\delta_{pred}$ and $\delta_{mit}$. This quantifies that time fluctuation plays a major role in the observed spread of the error-mitigated observable.  A similar analysis applies for the optimized and averaged noise channels in Figs.~\ref{fig:mit}e and~f, albeit with a distribution that is packed more closely around the origin. The tighter histograms in the inset of Fig.~\ref{fig:mit}e and~f highlight that both optimized and averaged noise channels effectively stabilize the temporal fluctuation of the error mitigation performance. In addition to the $Z$ parity analyzed here, we extend the comparison to observables of all weights in Supplementary Fig.~\ref{fig:allweights}. 

We note that additional sources of bias in the error-mitigated observables may remain, even for the average and optimized experiments. One source of bias that is important to consider, particularly for the averaged noise-case, is the effect of quasi-static noise on learning and mitigation, as detailed in Supplementary~\ref{app:theory}. For instance, quasi-static noise can lead to noise learning circuit fidelities that do not follow a clean single exponential decay with increasing depth, introducing bias in the mitigation that relies on an assumption of exponential decay. These effects are also relevant in the absence of any modulation, due to the natural temporal fluctuations in the TLS landscape and data collection over long periods of time. This essentially implies that the noise channel is, in practice, always quasi-static to some degree. As such, one remaining question of broad interest is how the quasi-static nature of noise, resulting from modulation at a shot-to-shot basis or intrinsic fluctuations, affects observable estimates and error mitigation. We explore this partially in Supplementary~\ref{app:theory}, but general investigation of these other sources and further closing the gap between ideal and mitigated observables remains a question for future work.
\\
\section{Discussion and Conclusion}

We experimentally demonstrate that noise in superconducting quantum processors can be stabilized by modulating the qubit-TLS interaction to improve the performance of error mitigation. Amongst the considered modes of operation, the optimized strategy gives the best performance, but remains exposed to random temporal fluctuations of qubit-TLS interaction between parameter re-optimizations, that could be particularly frequent at large qubit counts. By contrast, the averaged strategy smooths out small fluctuations and produces a more stable device noise model, albeit at the cost of a slightly increased sampling overhead for error mitigation, and potential bias in observable estimates. While these modulation strategies are complementary to the continued development of cleaner devices and novel designs to reduce the density and impact of defect two-level systems, we expect that the methods discussed in this work will be crucial for the reliability of error-mitigated quantum computation with solid-state processors, particularly at scales beyond exact verification.

\section*{Acknowledgements}
We acknowledge Dan Rugar, Robert M. Shelby, John Mamin, Jerry Tersoff, Sami Rosenblatt, Majo Lozano, Petar Jurcevic, Matthias Steffen and Oliver Dial for insightful discussions; Oliver Dial, Edward H. Chen, and Alireza Seif for feedback on the manuscript. For modeling the learning procedure under quasi-static noise, LCGG acknowledges support from the Army Research Office under Grant Number W911NF-21-1-0002. The views and conclusions contained in this document are those of the authors and should not be interpreted as representing the official policies, either expressed or implied, of the Army Research Office or the U.S. Government. The U.S. Government is authorized to reproduce and distribute reprints for Government purposes notwithstanding any copyright notation herein.

\section*{Data Availability}
Data available upon request to contributing authors.

\bibliography{reference}

\clearpage
\newpage

\setcounter{equation}{0}
\setcounter{figure}{0}
\setcounter{table}{0}
\setcounter{page}{1}
\setcounter{section}{0}
\makeatletter
\renewcommand{\theequation}{S\arabic{equation}}
\renewcommand{\thefigure}{S\arabic{figure}}
\renewcommand{\thetable}{S\arabic{table}}
\renewcommand{\thesection}{S\Roman{section}}
\renewcommand{\citenumfont}[1]{S#1}

\section*{Supplementary materials}

\section{optimized and averaged noise channel}
\subsection{Optimized noise channel} \label{sec:optnoise}
One way to improve qubit coherence is to actively monitor the temporal snapshot of the TLS landscape and choose $k_{TLS}$ that produces the best qubit property of interest.  
Assuming that we sample a TLS environment $i$, with $i\in \{1,2,\cdots, N_t\}$ for $N_t$ different TLS environments with corresponding $k_{TLS}$ values, the ideal unitary $\mathcal{U}$ is then exposed to a given TLS environment. Along with other noise sources, the given qubit-TLS interaction incurs a noise channel $\mathcal{E}_i$, resulting in a noisy operator $\tilde{\mathcal{U}}_i=\mathcal{U}\circ\mathcal{E}_i$. The observable expectation value is then computed as
\begin{equation} \label{eq:Oi}
    \braket{O}_i=\tr\left[ O\tilde{\mathcal{U}}_i(\rho) \right]
\end{equation} 
for a given initial state $\rho$. One may sample an optimal TLS environment $i$ that minimizes the bias of Eq.~\eqref{eq:Oi} from its ideal value. In practice, we choose a TLS control parameter that optimizes a representative metric, such as $T_1$ or a single-qubit randomized benchmarking result. 
This requires active monitoring of the TLS environment through via this representative metric. Throughout our study, we choose $\mathcal{P}_e$ (a proxy of $T_1$) as an optimization metric. We aim for the best target operation assuming the best coherence can be achieved by minimizing the qubit-TLS interaction. However, the qubit is still exposed to random fluctuations in qubit-TLS interaction between monitoring events.

\subsection{Averaged noise channel} \label{sec:averagednoise}
Alternatively, we may obtain an observable by averaging over randomly sampled TLS environments for each single shot measurement. In the limit of infinite shots, and assuming the TLS environment is static for the duration of a shot, the observable expectation value obtained by averaging over sampled TLS environments is given by
\begin{equation} \label{eq:Obar}
\begin{split}
   \Bar{\braket{O}} &= \sum_{i}p_i\braket{O}_i=\sum_{i} p_i\tr\left[ O\tilde{\mathcal{U}}_i(\rho) \right], \\
   &= \tr\left[ O \sum_{i}p_i\tilde{\mathcal{U}}_i(\rho) \right]
   =\tr\left[ O \bar{\mathcal{U}}(\rho) \right],
\end{split}
\end{equation}  
where $p_i$ is the probability of having noise realization $i$ at a given shot, and we define an averaged noise channel
\begin{equation} \label{eq:lambdabar}
    \bar{\mathcal{U}}=\sum_{i}p_i\tilde{\mathcal{U}}_i.
\end{equation}
The averaged noise channel may not be optimal in terms of minimizing the bias on the observable of interest. Nevertheless, the resulting variation in the observable value due to a local fluctuation of the TLS environment is reduced by averaging over the sampled TLS environments. The method aims to provide a stable observable estimate in the presence of natural fluctuations of the TLS environment on timescales shorter than the total duration of data collection for an experiment. This method only requires passive sampling of TLS environment from shot to shot, thus constant monitoring activities are not required, unlike the optimized noise channel scenario.  

\subsection{$T_1$ measurements of the optimized noise channels}

Figure~\ref{fig:t1}c illustrates the temporal fluctuation of qubit-TLS interactions. Strong qubit-TLS interactions are indicated by a dip in $\mathcal{P}_e$ (dark green colors), and their specific locations vary over time. The optimal $k_{TLS}$ is selected by finding a maximum $\mathcal{P}_e$ at each experiment, as indicated by a red cross. The black line is $k_{TLS}$ for the control experiment, where it is set to a fixed value. Figure~\ref{fig:t1}a shows the comparison between optimized $T_1$ (red triangles) and $T_1$ from the control experiment (black rectangles). Note that the optimal point selected based on Fig.~\ref{fig:t1}c changes over time due to temporal changes in the qubit-TLS interaction. The result shows a clear benefit from this optimization in terms of an improved overall $T_1$. 

\subsection{$T_1$ measurements of the averaged noise channels} \label{supp:T1avg}

To illustrate the impact of the averaged noise channel on qubit coherence, we measure the same qubit for the value of $k_{TLS}$ used in the control experiment (indicated as a black line in Fig.~\ref{fig:t1}c). To modulate the TLS landscape from shot to shot, we apply a slowly varying sine wave with a frequency of $1$Hz and an amplitude of $\pm 0.5$ for Fig.~\ref{fig:t1}, and triangular wave with an amplitude of $\pm0.2$ for Figs.~\ref{fig:model} and~\ref{fig:mit} in arbitrary unit of $k_{TLS}$. The modulation form is arbitrary chosen and we did not observe any obvious differences between sine and triangular wave modulation. The experiment is repeated for each shot at a $1$kHz rate, such that the slowly varying modulation effectively samples different quasi-static TLS environments for each shot. We then average $T_1$ data over various qubit-TLS interaction landscapes collected from 300 single shot measurements. Figure~\ref{fig:t1}a illustrates $T_1$ using time-varying modulation of the TLS control (blue circles). The averaged $T_1$ value is more stable than $T_1$ of the control experiment. However, the value remains stable without any additional monitoring or optimization.

\begin{figure}[t!]
    \centering
    \includegraphics[width=1.0\columnwidth]{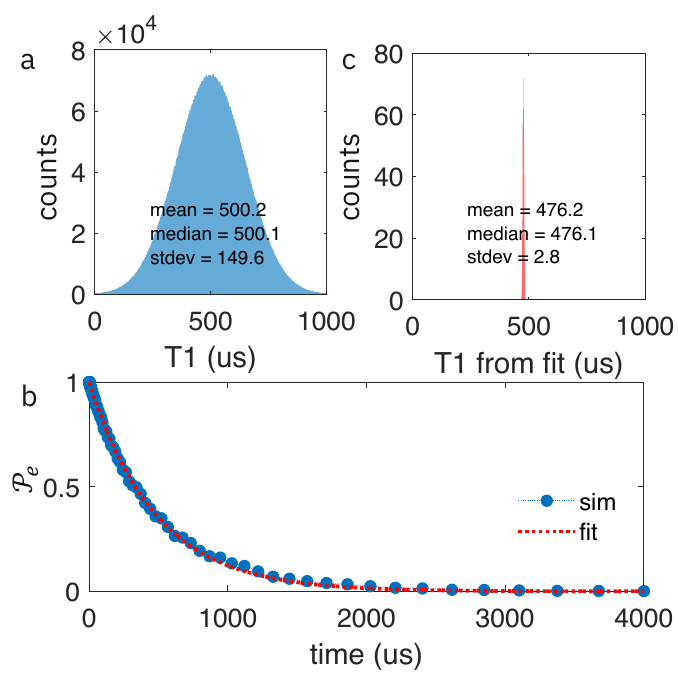}
    \caption{Simulations of $T_1$ with varying TLS control parameters. (a) We assume that the TLS states accessible by tuning the control parameter result in $T_1$ values that are normally distributed with a mean $T_1$ of 500 $\mu$s and a standard deviation of 150 $\mu$s. (b) $\mathcal{P}_e$ curve constructed by randomly selecting a static $T_1$ for a given sequence of X gates followed by delays, and a final measurement. We repeat this process 1,000 times. At each delay, the resulting set of $\mathcal{P}_e$ values is averaged together. The points are then fit with a decaying exponential to extract the averaged noise channel $T_1$ value. (c) Histogram of averaged noise channel $T_1$ values extracted from repeated simulations of (b).}
    \label{fig:t1supp2}
\end{figure}

Here we provide a simple sketch and numerical calculation to explain how varying a TLS control parameter at a suitable rate can be used to stabilize the $T_1$ values over time. 
If the TLS states that are accessible by spectral diffusion are the same as, or substantially overlap with, the TLS states that are accessible by varying the TLS control parameter, then we can use the TLS control parameter to average out the time variation in $T_1$. Here, we vary the TLS control parameter at a rate that is fast compared to the entire experiments but slow compared to the single shot execution time. This way, every repetition of the experiment experiences a different quasi-static TLS configuration, and therefore a different $T_1$. 
Assuming we can sample from the full $T_1$ distribution as illustrated in Fig.~\ref{fig:t1supp2}a, we average the results. We then construct the probability of observing $\ket{1}$ state as shown in Fig.~\ref{fig:t1supp2}b. When we extract the $T_1$ parameter from the decay curve, we find that the decay rate is similar to the mean value of the $T_1$ distribution, but whose exact value depends on the details of the original distribution. One thing that should be clear from the above, is that the values extracted and plotted in Fig.~\ref{fig:t1supp2}c are far lower than the maximum values one could observe from sampling the $T_1$ distribution, as the averaging procedure naturally includes relatively strong qubit-TLS interactions.

\begin{figure}[ht]
    \centering
    \includegraphics[width=1.0\columnwidth]{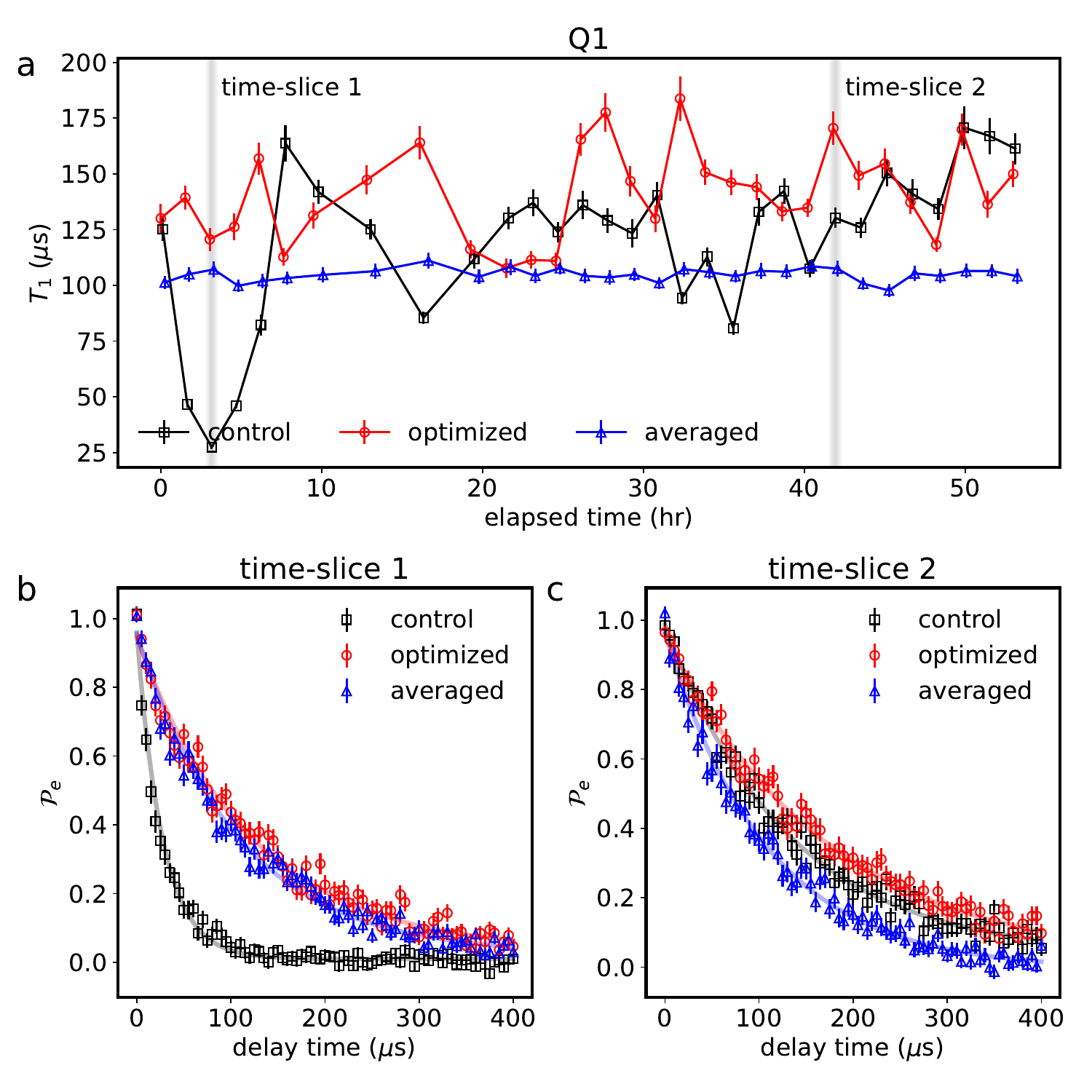}
    \caption{(a) Measured $T_1$ values for the control, optimized, and averaged noise strategy for $Q_1$. Two shaded regions separately illustrate the $T_1$ decay curve, highlighting fluctuating properties of control experiment where it shows minimum $T_1$ at (b) and recovered $T_1$ value at (c).}
    \label{fig:t1supp}
\end{figure}

\begin{figure}[ht]
    \centering
    \includegraphics[width=1.0\columnwidth]{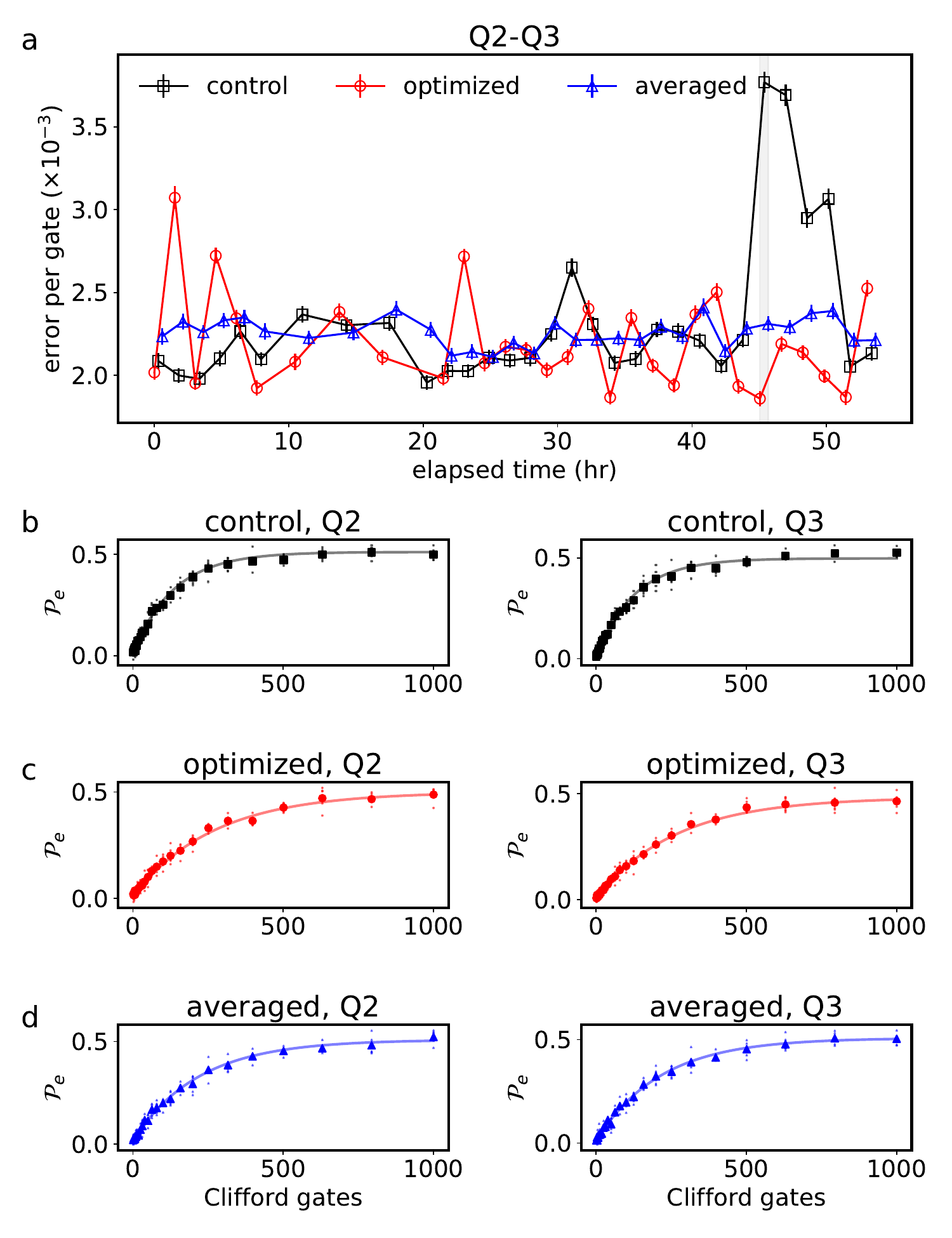}
    \caption{(a) Measured error per gate (EPG) values for the control, optimized, and averaged noise strategy for $Q2-Q3$. A shaded region illustrates the two-qubit randomized benchmarking decay curve, highlighting fluctuating properties of control experiment. (b-d) show detailed decay curves for control, optimized, and averaged noise strategy, respectively, at the time point highlighted as a gray vertical line in (a). Each point in (b-d) is averaged over six different randomized benchmark circuit realizations (small symbols).}
    \label{fig:tqrbsupp}
\end{figure}

Figure~\ref{fig:t1supp} shows explicit examples of the full T1 decays, with single-exponential fits to $\mathcal{P}_e$ for $Q1$ on the device with averaged noise, at two different time points.

\section{device property}
\label{app:device}

In this section, we provide device properties characterized over $\sim50$ hours for six qubits. The characterization is carried out to provide general properties of the device beyond $T_1$ measurements.
Measurements are done at a separate time from experiments in the main text.
The same set of measurements is performed for three different cases; control, optimized, and averaged noise strategies (see main text for details). For the optimized noise strategy, we carry out $\mathcal{P}_e$ based optimization for all six qubits before measuring each property, and averaged noise strategy uses the same condition described in Figs.~\ref{fig:model} and~\ref{fig:mit}. 

Table~S1 summarizes single-qubit properties $T_1$, $T_2$, and the error per gate (EPG) for single-qubit gates. Here, $T_2$ is measured with an echo $\pi$-pulse placed in the middle of the delay time and EPG is obtained using a randomized benchmarking protocol~\cite{Knill2008, Magesan2011}. Likewise, table~S2 illustrates EPG for two-qubit gates. Here, the randomized benchmarking protocol for two-qubit gates is performed simultaneously on qubit pairs $Q1$--$Q2$, $Q3$--$Q4$, and $Q5$--$Q6$, and likewise for pairs $Q2$--$Q3$ and $Q4$--$Q5$. Figure~\ref{fig:tqrbsupp} shows explicit examples of the full decays of two-qubit randomized benchmarking results on $Q2$--$Q3$, with single-exponential fits to excitation probability $\mathcal{P}_e$ for $Q2$ and $Q3$ on the device.

\begin{table*}[!ht]
\centering
\resizebox{\linewidth}{!}{%
\begin{tabular}{||c | c | c c c c| c c c c| c c c c||} 
 \hline
 & & & \multicolumn{2}{ c  }{control}&  &  & \multicolumn{2}{ c  }{optimized} & & & \multicolumn{2}{ c }{averaged}  &\\
  qubit & property & median & mean & max & min & median & mean & max & min & median & mean & max & min \\ [0.5ex] 
 \hline\hline
 $Q_1$ & $T_1$ ($\mu s$) & 129.14 & 120.23 $\pm$ 34.55 & 170.76 & 27.44 & 139.36 & 141.07 $\pm$ 20.25 & 183.75 & 107.69 & 104.89 & 104.82 $\pm$ 2.78 & 111.09 & 97.72 \\
 & $T_2$ ($\mu s$) & 156.72 & 143.84 $\pm$ 43.90 & 224.50 & 40.43 & 167.16 & 174.70 $\pm$ 28.59 & 239.27 & 121.40 & 133.45 & 131.66 $\pm$ 10.19 & 152.95 & 114.26 \\
 & $EPG$ ($ \times 10^{-4}$) & 1.96 & 2.38 $\pm$ 1.11 & 6.54 & 1.34 & 1.76 & 1.75 $\pm$ 0.26 & 2.31 & 1.12 & 2.16 & 2.16 $\pm$ 0.17 & 2.44 & 1.81 \\
 \hline
$Q_2$ & $T_1$ ($\mu s$) & 132.06 & 126.08 $\pm$ 25.43 & 172.66 & 68.79 & 137.36 & 134.05 $\pm$ 21.86 & 174.76 & 87.86 & 118.65 & 117.78 $\pm$ 7.84 & 131.37 & 88.23 \\
 & $T_2$ ($\mu s$) & 158.89 & 157.73 $\pm$ 30.68 & 210.31 & 102.96 & 164.40 & 162.17 $\pm$ 29.35 & 223.89 & 94.33 & 140.19 & 141.00 $\pm$ 12.61 & 172.32 & 106.52 \\
 & $EPG$ ($ \times 10^{-4}$) & 1.82 & 1.82 $\pm$ 0.25 & 2.30 & 1.37 & 1.86 & 1.87 $\pm$ 0.27 & 2.47 & 1.34 & 2.06 & 2.09 $\pm$ 0.25 & 2.59 & 1.50 \\
 \hline
$Q_3$ & $T_1$ ($\mu s$) & 100.26 & 103.12 $\pm$ 16.99 & 137.07 & 76.25 & 114.40 & 114.02 $\pm$ 18.54 & 150.75 & 68.89 & 80.60 & 82.14 $\pm$ 5.12 & 97.02 & 75.47 \\
 & $T_2$ ($\mu s$) & 134.46 & 137.77 $\pm$ 26.31 & 209.13 & 97.90 & 158.68 & 155.89 $\pm$ 28.59 & 237.64 & 89.42 & 105.81 & 107.19 $\pm$ 9.79 & 131.03 & 91.17 \\
 & $EPG$ ($ \times 10^{-4}$) & 2.23 & 2.67 $\pm$ 1.21 & 7.27 & 1.82 & 2.08 & 2.17 $\pm$ 0.52 & 4.43 & 1.45 & 2.75 & 2.78 $\pm$ 0.24 & 3.32 & 2.31 \\
 \hline
$Q_4$ & $T_1$ ($\mu s$) & 108.73 & 107.83 $\pm$ 20.40 & 141.77 & 32.89 & 109.82 & 110.27 $\pm$ 15.72 & 146.51 & 74.81 & 100.07 & 99.76 $\pm$ 4.45 & 111.37 & 91.90 \\
 & $T_2$ ($\mu s$) & 133.28 & 130.46 $\pm$ 25.43 & 163.15 & 46.67 & 139.86 & 139.18 $\pm$ 21.54 & 180.32 & 88.17 & 124.24 & 125.50 $\pm$ 9.70 & 149.02 & 109.27 \\
 & $EPG$ ($ \times 10^{-4}$) & 2.52 & 2.98 $\pm$ 1.34 & 7.13 & 1.53 & 2.15 & 2.44 $\pm$ 0.86 & 4.97 & 1.63 & 2.36 & 2.39 $\pm$ 0.41 & 3.81 & 1.81 \\
 \hline
$Q_5$ & $T_1$ ($\mu s$) & 110.62 & 98.34 $\pm$ 40.19 & 159.81 & 16.75 & 82.50 & 86.67 $\pm$ 41.04 & 157.68 & 12.92 & 83.76 & 86.50 $\pm$ 26.83 & 143.45 & 43.04 \\
 & $T_2$ ($\mu s$) & 117.40 & 100.94 $\pm$ 43.63 & 172.05 & 17.65 & 84.12 & 91.16 $\pm$ 47.57 & 175.89 & 12.00 & 81.07 & 90.47 $\pm$ 31.02 & 161.07 & 34.97 \\
 & $EPG$ ($ \times 10^{-4}$) & 2.75 & 4.94 $\pm$ 4.10 & 14.63 & 1.58 & 3.80 & 5.93 $\pm$ 4.37 & 18.89 & 2.00 & 4.67 & 4.88 $\pm$ 2.31 & 9.40 & 1.71 \\
 \hline
$Q_6$ & $T_1$ ($\mu s$) & 61.13 & 60.37 $\pm$ 7.49 & 74.08 & 40.02 & 67.22 & 66.21 $\pm$ 6.26 & 76.64 & 48.53 & 55.12 & 55.63 $\pm$ 2.36 & 59.98 & 47.39 \\
 & $T_2$ ($\mu s$) & 79.10 & 77.57 $\pm$ 10.32 & 97.57 & 54.33 & 86.40 & 84.15 $\pm$ 8.86 & 107.09 & 60.50 & 71.60 & 71.51 $\pm$ 4.93 & 82.68 & 60.30 \\
 & $EPG$ ($ \times 10^{-4}$) & 3.25 & 3.39 $\pm$ 0.41 & 4.62 & 2.91 & 3.07 & 3.32 $\pm$ 0.68 & 5.13 & 2.62 & 4.26 & 4.23 $\pm$ 0.49 & 6.31 & 3.14 \\
 \hline
 \end{tabular}%
 }
 \label{table:1Q}
\caption{{\bf{Summary of single qubit properties}}. Reported $T_1$ and $T_2$ values are obtained from $31$ points measured over $\sim50$ hours.}

\end{table*}

\begin{table*}[!ht]
\centering
\begin{tabular}{||c | c | c c c c | c c c c | c c c c||} 
 \hline
 & & &\multicolumn{2}{ c }{control}& & &\multicolumn{2}{ c }{optimized}& & &\multicolumn{2}{ c }{averaged}&\\
  qubit & property & median & mean & max & min & median & mean & max & min & median & mean & max & min \\ [0.5ex] 
 \hline\hline
 Q1-Q2 & $EPG$ ($ \times 10^{-3}$) & 2.78 & 2.91 $\pm$ 0.48 & 4.61 & 2.50 & 2.58 & 2.61 $\pm$ 0.19 & 3.04 & 2.29 & 3.04 & 3.04 $\pm$ 0.12 & 3.34 & 2.82 \\
 \hline
Q2-Q3 & $EPG$ ($ \times 10^{-3}$) & 2.13 & 2.32 $\pm$ 0.45 & 3.77 & 1.96 & 2.11 & 2.19 $\pm$ 0.29 & 3.07 & 1.86 & 2.26 & 2.26 $\pm$ 0.08 & 2.41 & 2.11 \\
 \hline
Q3-Q4 & $EPG$ ($ \times 10^{-3}$) & 4.25 & 4.41 $\pm$ 0.89 & 6.87 & 3.36 & 3.77 & 3.82 $\pm$ 0.28 & 4.55 & 3.41 & 4.34 & 4.29 $\pm$ 0.28 & 4.73 & 3.79 \\
 \hline
Q4-Q5 & $EPG$ ($ \times 10^{-3}$) & 5.74 & 6.05 $\pm$ 2.64 & 15.57 & 2.96 & 4.66 & 5.11 $\pm$ 1.42 & 8.11 & 2.97 & 5.24 & 5.43 $\pm$ 1.55 & 8.55 & 3.20 \\
 \hline
Q5-Q6 & $EPG$ ($ \times 10^{-3}$) & 5.58 & 5.91 $\pm$ 0.99 & 8.13 & 4.40 & 6.05 & 6.00 $\pm$ 1.04 & 8.58 & 4.04 & 5.99 & 6.03 $\pm$ 0.78 & 7.69 & 4.42 \\
 \hline
 \end{tabular}
 \label{table:2Q}
\caption{{\bf{Summary of two-qubit gate infidelity}}. Reported two-qubit infidelities are obtained from $31$ points measured over $\sim50$ hours.}
\end{table*}

\section{Characterizing the model parameters} \label{app:getlambda}

By applying Pauli twirling, we can assume the noise channel to be a Pauli channel; that is $\mathcal{E}_i(\rho) = \sum_{k} \alpha_{k,i} P_k(\rho)P_k^{\dag}$ where the $\alpha_{k,i} \geq 0$ terms sum up to one. Sampling TLS environment $i$ by changing modulation parameter $k_{TLS}$ changes the noise channel $\mathcal{E}_{i}$, and, consequently, the corresponding effective gate operation $\tilde{\mathcal{U}}_{i}$. In this section we study how the different TLS modulation strategies affect the noise associated with two layers of concurrent CZ gates.

Learning a full $n$-qubit Pauli noise channel requires the characterization of $4^n$ parameters $\alpha_k$, and is therefore feasible only for a small number of qubits. The sparse Pauli-Lindblad noise model proposed in~\cite{berg2023probabilistic} provides a scalable alternative obtained by imposing more structure on the noise. In particular, the noise is assumed to be of the form $\mathcal{E}(\rho) =\exp[\mathcal{L}](\rho)$, where $\mathcal{L}$ represents a Lindbladian
\[
\mathcal{L}(\rho) = \sum_{k\in\mathcal{K}}\lambda_k (P_k\rho
P_k^{\dag} - \rho),
\]
 with Pauli jump terms and non-negative coefficients $\lambda_k$.  Sparsity is achieved by making the reasonable assumption that noise originates locally on individual or connected pairs of qubits, which is done by restricting the set of generators $\mathcal{K}$ to all one- and two-local Pauli terms in accordance with the qubit topology. 
The set of model parameters $\lambda_k$ are characterized by measuring the set of Pauli fidelities. The Pauli fidelity for a Pauli $P_k$ is given by 
\begin{equation}\label{eq:FidelityLambda}
f_k = \frac{1}{2^n}\tr(P_k\mathcal{E}(P_k)) =
\exp\Big(-2\sum_{k'\in\mathcal{K}}\lambda_{k'}\langle
k,k'\rangle_{\mathrm{sp}}\Big),
\end{equation}
where the symplectic inner product $\langle k,k'\rangle_{\mathrm{sp}}$ is zero if Paulis $P_k$ and $P_{k'}$ commute, and one otherwise.  As described in more detail in~\cite{berg2023probabilistic} \textcolor{black}{and references therein}, repeated application of the noisy layer of gates allows us to learn fidelity pairs $f_kf_{k'}$, where $P_{k'} = \pm \mathcal{U}(P_k)$. In general, it is not possible to measure the individual fidelities in a readout-error free manner~\cite{chen2023learnability}, and a symmetry assumption that equates the fidelities appearing in each pair is therefore imposed.  Given the fidelities, a nonnegative least-squares problem based on Eq.~\eqref{eq:FidelityLambda} can then be used to obtain the model parameters $\lambda$.

The sparse Pauli-Lindblad noise model can equivalently be expressed as a series of Pauli channels of the form $\Lambda_k(\rho) = (w_k \rho + (1-w_k)P_k\rho P_k^{\dag})$, where $w_k = \frac{1}{2}(1 + \exp(-2\lambda_k))$. The fact that these Pauli channels commute makes the noise model ideal for probabilistic error cancellation~\cite{temme2017error,berg2023probabilistic}, since each channel can be inverted independently. Here, the inverse channel is non-physical and the observable is therefore reconstructed through post-processing. Once we insert the appropriate canceling Paulis following the quasi-probability described by $w_k$, the measured outcome is multiplied by a pre-factor given by
\begin{align}
    \gamma = \prod_{k\in\mathcal{K}}(2w_k - 1)^{-1} =
\exp\Big(\sum_{k\in\mathcal{K}} 2\lambda_k\Big).
\end{align}
While we obtain an unbiased estimate, variance is amplified by this pre-factor. Therefore, more sampling is required to compensate the increased variance due to the scaling. For example, the sampling overhead is proportional to $\gamma^2$ for PEC~\cite{berg2023probabilistic}, thus we refer $\gamma$ to as a \emph{sampling overhead}. The sampling overhead is a useful metric connecting the underlying noise strength to an experimental overhead for error mitigation. \textcolor{black}{For instance, we track the sampling overhead to visualize the overall model coefficient fluctuation in Fig.~\ref{fig:model}e. The sampling overhead also provides a useful estimation on the relative runtime cost between different error rates. For instance, the worst (best) case sampling overhead for a given set of two qubit layers in Fig.~\ref{fig:model}e is roughly $\gamma_w=1.13$ ($\gamma_b=1.06$). The ratio $(\gamma_w^2/\gamma_b^2)^N$ could be translated as a relative extra sampling cost for PEC for the benchmark circuit with a repeated unit layer depth $N$ between the worst and the best error rate scenario. As an example, the worst error rate scenario requires $\sim13$ ($\sim167$) times more circuits to perform PEC than the best error rate scenario for $N=20$ ($N=40$).}

The learning protocol consists of two steps. First the learning procedure extracts a set of Pauli fidelities, $\{f_b\}_{b\in\mathcal{B}}$, defined by a set of Pauli operators $\mathcal{B}$ much smaller than the set of all Pauli operators. Second, we use nonnegative least-squares minimization to extract the parameters $\{\lambda_k\}_{k\in{\mathcal{K}}}$ of our sparse Pauli-Lindblad model. The Pauli fidelities that are used to extract model parameters described in Fig.~\ref{fig:model} are obtained by repeating the gate sequences an even numbers of times ($n=0,4,12,24,64$) for both layer 1 and layer 2, independently. At each depth, we generate $60$ circuit instances with randomly sampled Pauli for Pauli twirling for gates and final measurements. All experiments were carried out with $1$kHz repetition rate and observable values for each circuit are estimated from 32 single shot measurements.

\section{Quasi-static noise theory} \label{app:theory}
We assume that our quasi-static noise is defined by a family of Pauli channels $\{\mathcal E_i\}$.
Here, we examine the impact of quasi-static noise on the noise learning procedure. The noise learning protocol consists of estimating Pauli fidelities by $d$-fold repetition of noisy identity operations, from which we deduce the generator coefficients of the assumed noise model. When it comes to the quasi-static noise model, the generator coefficients are no longer represented by a single value but by an ensemble that follows a specific probability distribution. The conventional noise learning protocol may induce inaccuracy in estimating (i) the Pauli fidelities (see section~\ref{sec:learning}), and (ii) the ensemble average of the generator coefficients (see Sec.~\ref{app:issue-gen}). As a result, the averaged channel may introduce an additional bias error that cannot be corrected with PEC during the mitigation process. 

For quasi-static error we will show in this section that circuit fidelities no longer decay exponentially as a function of depth but will generically have a complicated dependence on moments of the error distribution up to the given depth. This has a profound impact on our error mitigation schemes, where one first uses the learning data under the assumption of exponential decay to fit an error model that is then used for mitigation. Concretely, this mismatch between assumed and actual decay shape induces a bias error that cannot be corrected with increased sampling. This induced bias is analogous to that intrinsic to extrapolation based error mitigation protocols when the functional form of the error extrapolation is not known \textit{a priori}. Moreover, quasi-static noise strengthens the dependence of mitigation performance on the learning procedure, as the magnitude of the bias error can depend on the specific depths chosen in the learning sequence.
As such, it represent one of the error sources along with other hardware non-idealities. We expand upon these points in the following subsections with supporting analytical calculations and numerical simulations.

\subsection{Pauli fidelity estimation with a quasi-static noise channel} \label{sec:learning}

To obtain the noise model parameters $\lambda$, we first experimentally obtain the noise channel Pauli fidelities. In reality, the error rates of two-qubit gates are an order of magnitude higher than those of single-qubit gates. Therefore, we focus on extracting the Pauli fidelities of the two-qubit gates assuming that single-qubit gates are ideal.   
The learning procedure consists of even-number repetitions of \gates{CZ} gates, which ideally implements an identity operation. Two noisy \gates{CZ} gates represent a noisy identity operator, which can be expressed as $\tilde{\mathcal{U}}^{(I)}_i=\tilde{\mathcal{U}}^{(CZ)}_i\circ \tilde{\mathcal{U}}^{(CZ)}_i$. Here, the noisy operation is described as $\tilde{\mathcal{U}}^{(CZ)}_i=\mathcal{U}^{(CZ)}\circ\mathcal{E}_i^{(CZ)}$, where $\mathcal{U}^{(CZ)}$ is an ideal \gates{CZ} operation and $\mathcal{E}_i^{(CZ)}$ is the Pauli noise channel incurred during the \gates{CZ} operation with a specific TLS environment $i$.

Consider first a fixed TLS environment. Following Eq.~\eqref{eq:FidelityLambda}, the Pauli noise channel of the net-identity gate is then characterized by the fidelities of the corresponding two-qubit Paulis $P_k$ and $P_{k'} = \mathcal{U}^{(CZ)}(P_k)$ as
\begin{align} 
     &\nonumber\frac{1}{2^n} \tr\left[ P_k^\dagger \tilde{\mathcal{U}}^{(I)}_i(P_k) \right] = \frac{1}{2^n} \tr\left[ P_k^\dagger \tilde{\mathcal{U}}^{(CZ)}_i\circ \tilde{\mathcal{U}}^{(CZ)}_i(P_k) \right] \\ 
     &= \frac{1}{2^n} \tr\left[ P_k^\dagger \tilde{\mathcal{U}}^{(CZ)}_i(P_{k'})f_{k,i}\right] = f_{k,i}f_{k',i}, \label{eq:fid}
\end{align}
where $f_{k,i}=(1/2^n)\tr\left[ P_k^\dagger \mathcal{E}_{i}^{(CZ)} P_k \right]$ is the Pauli fidelity for $P_{k}$ of the noise channel. The fidelity estimate of the noisy identity operator comes as the product of two Pauli fidelities, which may themselves be indistinguishable~\cite{berg2023probabilistic,chen2023learnability}. The fidelity is then extracted by repeating the noisy identity operation $d$ times (equivalently applying the \gates{CZ} gate $2d$ times), 
\begin{equation} \label{eq:drep}
    \frac{1}{2^n}\tr\left[ P_a^\dagger \tilde{\mathcal{U}}^{(I)}_i\circ\tilde{\mathcal{U}}^{(I)}_i\cdots\circ \tilde{\mathcal{U}}^{(I)}_i(P_a) \right] = ( f_{k,i}f_{k',i} )^d,
\end{equation}
where we extract the fidelity $f_{k,i}f_{k',i}$ from the obtained decay curve that is isolated from SPAM error~\cite{berg2023probabilistic}.

Now we consider a case with an averaged noise channel. In this case, we assume that the noise channel is an ensemble of the individually sampled noise channels from a given TLS environment distribution as discussed in Sec.~\ref{sec:averagednoise}. We first define an averaged fidelity as 
\begin{equation}
    \begin{split}
        \bar{f}_k\bar{f}_{k'}&=\mathbb{E}[ \hat{f}_k\hat{f}_{k'} ]=\sum_i p_i f_{k,i}f_{k',i} \\
        &= \sum_i p_i \frac{1}{2^n}\tr\left[ P_k^\dagger \tilde{\mathcal{U}}^{(I)}_i(P_k) \right] \\
        &= \frac{1}{2^n}\tr\left[ P_k^\dagger \sum_i p_i\tilde{\mathcal{U}}^{(I)}_i(P_k) \right] \\
        &= \frac{1}{2^n}\tr\left[ P_k^\dagger \bar{\mathcal{U}}^{(I)}(P_k) \right], \\
    \end{split}
\end{equation}
where $\mathbb{E}[\hat{f}]$ is a statistical expectation value of the random variable fidelity $\{\hat{f}\}$, and the averaged noise channel of the net-identity gate, $\bar{\mathcal{U}}^{(I)}$, is defined analogously to that of the individual CZ gate as in Eq.~\eqref{eq:lambdabar}. Note that because $P_k$ and $P_{k}'$ are indistinguishable we cannot uniquely define an averaged noise channel for a single CZ gate with this procedure, unless we use a symmetry assumption to break $\bar{\mathcal{U}}^{(I)}$ into the product of two averaged CZ gates.

However, our learning procedure involves $2d$ repetitions of \gates{CZ} gates, and given the rate of modulation used for our average channel experiments, we assume that each sampled noise channel is \emph{quasi-static} such that the noise realization does not change during a single instance of the circuit. Thus, over the duration of all $2d$ repeated applications of the noisy unitary the noise channel is the same. In other words, the noise channel changes its configuration from shot-to-shot, but is constant within a shot. In this case, the fidelity obtained after $2d$ repetitions of the \gates{CZ} gate becomes
\begin{equation} \label{eq:repeat}
    \begin{split}
    & \sum_i p_i \frac{1}{2^n}\tr\left[ P_k^\dagger \tilde{\mathcal{U}}^{(I)}_i\circ\tilde{\mathcal{U}}^{(I)}_i\cdots\circ \tilde{\mathcal{U}}^{(I)}_i(P_k) \right] \\ 
    & =\sum_i p_i \left( f_{k,i}f_{k',i} \right)^d \\
    & = \mathbb{E}[ (\hat{f}_k\hat{f}_{k'})^d ].
    \end{split}
\end{equation}
Note that this is not identical to $2d$ repetitions of the averaged fidelity, namely, $\mathbb{E}[ (\hat{f}_k\hat{f}_{k'})^d ] \neq  (\mathbb{E}[ \hat{f}_k\hat{f}_{k'} ])^d = (\bar{f}_a\bar{f}_b)^d$ in general, but instead will depend on moments of the distribution up to order $d$. More generally, we can define the $d$-fold averaged noisy gate 
\begin{equation}
    \bar{\mathcal U}^{(I)}_d = \sum_i p_i \mathop{\bigcirc}_{j=1}^d \tilde{\mathcal U}^{(I)}_i,
\end{equation}
for which $\bar{\mathcal U}_d\neq \mathop{\bigcirc}_{j=1}^d \bar{\mathcal U}_i$ for quasi-static noise. In other words, the average noise channel of $d$-fold noisy operations is not identical to $d$-fold applications of the average noise channel. This means that repeated application of the noisy unitary no longer amplifies errors as expected. The impact of quasi-static noise has previously been investigated for randomized benchmarking \cite{Ball16,Fong17,FR21}, where its impact is qualitatively similar to what we show here for Pauli learning. 

If the noise channel is randomly sampled for each noisy identity operation within the circuit of a single shot, and sampled noise channels are uncorrelated to each other, then the effective noisy operation is $\mathop{\bigcirc}_{j=1}^d \bar{\mathcal U}_i$, and the repeated learning process satisfies 
\begin{equation} 
    \begin{split}
        & \frac{1}{2^n}\tr\left[ P_k^\dagger \bar{\mathcal{U}}^{(I)}\circ\bar{\mathcal{U}}^{(I)}\cdots\circ \bar{\mathcal{U}^{(I)}}(P_k) \right] \\ 
    & =\left( \bar{f}_{k}\bar{f}_{k'} \right)^d \\
    & = (\mathbb{E}[ \hat{f}_k\hat{f}_{k'} ] )^d.
    \end{split} \label{eqn:circdecay}
\end{equation}
Under this full Markovian assumption, we obtain the averaged noise channel fidelity pair, $\bar{f}_k\bar{f}_{k'}$, in a SPAM free manner by fitting the data  from even depth repetitions of \gates{CZ} gates with an expected decaying function. Realizing the full Markovian assumption would require us to modulate $k_{TLS}$ at a rate comparable to the gate speed, and practical realization of such operation mode requires further considerations on hardware implementations. 
For this reason, our experiment is carried out with slow modulation whose description of the modulated noise channel is close to the quasi-static case.

In the quasi-static case, it is clear that, in general, $\mathbb{E}[\hat{f}_k^d\hat{f}_{k'}^d]\neq\mathbb{E}[\hat{f}_k\hat{f}_{k'}]^d$.
Nevertheless, the discrepancy could be small in certain regimes. To be more specific, we consider the case where the fluctuations around the average fidelities are an additive random variable with zero mean, i.e., $f_{k,i}=\bar{f}_{k}+\delta_{k,i}$, where $\delta_{k,i}$ describes the deviation of the Pauli fidelity from its averaged value and satisfies $\sum_i p_i\delta_{k,i}=0$. Then in the $\delta_{k(k'),i}\sim\delta\ll 1$ limit, we have
\begin{equation} \label{app:approx}
    \begin{split}
    &\mathbb{E}[ ( \hat{f}_k\hat{f}_{k'})^d ]  =\sum_i p_i \left( f_{k,i}f_{k',i} \right)^d \\ =& \sum_i p_i (\bar{f}_k+\delta_{k,i})^d(\bar{f}_{k'}+\delta_{k',i})^d \\
    =&\sum_i p_i 
        \left[ \left( \bar{f}_{k}\bar{f}_{k'} \right)^d
         + d\bar{f}_k^{d-1}\bar{f}_{k'}^d\delta_{k,i} + d\bar{f}_{k}^d\bar{f}_{k'}^{d-1}\delta_{k',i} +\mathcal{O}(\delta^2) \right] \\
    =& \left( \bar{f}_{k}\bar{f}_{k'} \right)^d
    +\mathcal{O}(\delta^2)
    \simeq [\mathbb{E}[ \hat{f}_k\hat{f}_{k'} ] ]^d.
    \end{split}
\end{equation}
Therefore, under the assumption that the fluctuations in the Pauli fidelities due to the changing TLS environment are weak and average to zero, we may still extract meaningful approximations of the averaged Pauli fidelities from our learning procedure.

While small additive changes to the Pauli fidelities due to quasi-static noise may not strongly impact our learning procedure, it is also instructive to consider multiplicative changes. A multiplicative change in $f_k$ can be modeled as an additive change in the rates of its generator. Writing $f_k = e^{-\Gamma_k}$ where $\Gamma_k$ is twice the sum of generator coefficients for all Pauli generators that anti-commute with $P_k$, we see that any deviation $\Gamma_k \rightarrow \Gamma_k + \delta$ naturally becomes $f_k \rightarrow e^{-\delta}f_k$. An additive change in the generator rate is the expected impact of changing $T_1$ with different TLS environments, and so this is the natural model of our experiments.
As a simple example, consider $\Gamma_k$ Gaussian distributed with mean $\mu_k$ and standard deviation $\sigma_k$. As a result, $f_k$ follows what is known as a log-normal distribution. Raising $f_k$ to power $d$ simply enhances the mean and standard deviation by a factor of $d$, and as $\left<f^d_k\right> = \left<e^{-d\Gamma_k}\right>$ is the the moment generating function of a Gaussian distribution evaluated at $-d$, we have that
\begin{align}
    \mathbb{E}\left[f^d_k\right] = \exp\left(-d\mu_k + \frac{1}{2}d^2\sigma_k^2\right).\label{eqn:curve_mult}
\end{align}
Note that even for $d=1$ this already differs from the simple ensemble average of the generator sum, or namely, $e^{-\left<\Gamma_k\right>} = e^{-\mu_k}\neq \braket{e^{-\Gamma_k}}=e^{-\mu_k+\sigma_k^2/2}$. As before, the impact of quasi-static noise appears at order $d^2$ and depends on the small parameter $\sigma^2_k$.

\subsection{Model parameter estimation with a quasi-static noise channel} \label{app:issue-gen}

The full learning procedure to obtain the parameters of the error model uses a non-negative constraint on the generator coefficients $\lambda_k$ that leaves the least-squares optimization without an analytic solution. 
To study the impact of quasi-static noise in isolation, we assume the infinite shot limit and that there is no SPAM error or other experimental imperfections, such that the non-negative constraint is no longer needed and we can analytically solve the least-squares minimization. 

For illustrative purposes, we consider a simpler problem than the full model learning, which is to estimate a single mean decay rate $\mu_k$ from the decay curve of a single Pauli fidelity $f_k$ (assumed non-degenerate). For a sequence of depths $\{d_j\}$ we minimize the cost function 
\begin{align}
    C = \sum_j\left|\log\left<f^{d_j}_k\right> - \tilde \mu_k d\right|^2
\end{align}
as a function of the parameter $\tilde\mu_k$, which is our estimate for a ground truth $\mu_k$ . The exact solution to this is given by
\begin{align}
    \nonumber\tilde\mu_k &= \frac{-\sum_j d_j\log\left<f^{d_j}_k\right>}{\sum_j d^2_j} = \frac{\sum_j d^2_j\mu_k - \frac{1}{2}d_j^3\sigma_k^2}{\sum_j d^2_j} \\ &= \mu_k -\frac{1}{2}\sigma_k^2\frac{\sum_j d^3_j}{\sum_j d^2_j} \equiv \mu_k -\frac{1}{2}\sigma_k^2d_{\rm eff},
\end{align}
where in the final expression we have defined an effective depth $d_{\rm eff}$. The key take-away from this result is that not only is $\tilde\mu_k \neq \mu_k$ but $\tilde\mu_k$ ($\tilde f_k$) contains an additive (multiplicative) error term that depends on the specific sequence of depths chosen during the learning procedure.

\subsection{Error mitigation with a quasi-static noise channel} \label{app:issue-mit}

Continuing our simple model with quasi-static Gaussian distributed gernerator rates, we evaluate how this impacts mitigation of a trivial circuit. We consider a circuit formed by repetitions of the noisy layer $d$ times, for which the expectation value of the Pauli $P_k$ is given by
\begin{align}
    \left<P_k\right> = \left<P_k\right>_{\rm ideal}\mathbb{E}\left[f^d_k\right] = \exp\left(-d\mu_k + \frac{1}{2}d^2\sigma_k^2\right)\left<P_k\right>_{\rm ideal}.
\end{align}
For this circuit, PEC in the infinite shot limit amounts to dividing the outcome by $\tilde f^d_k$ of the previous subsection, which gives
\begin{align}
    \nonumber\left<P_k\right>_{\rm mit} &= \frac{\left<P_k\right>}{\tilde f^d_k}= \left<P_k\right>_{\rm ideal}\frac{\mathbb{E}\left[f^d_k\right]}{\tilde f^d_k} \\ \nonumber &= \frac{\exp\left(-d\mu_k + \frac{1}{2}d^2\sigma_k^2\right)}{\exp\left(-d\mu_k +\frac{1}{2}dd_{\rm eff}\sigma_k^2\right)}\left<P_k\right>_{\rm ideal} \\ &= \exp\left(\frac{1}{2}\sigma^2_kd\left[d-d_{\rm eff}\right]\right)\left<P_k\right>_{\rm ideal}. \label{eqn:mitigated}
\end{align}
As this shows, even for this idealized toy model the mitigated value does not achieve its ideal value, as quasi-static noise induces a multiplicative bias error that scales exponentially with the variance of the noise distribution. \textcolor{black}{Therefore, choosing $k_{TLS}$ points that results in relatively uniform qubit characteristics is helpful in improving error mitigation with quasi-static noise channel.} 

However, for this simple case of Gaussian distributed quasi-static noise in the generator rates (e.g., $1/T_1$) this bias can be removed if the effective dimension of the learning sequence $d_{\rm eff}$ is chosen to match the depth of the target circuit. As quasi-static noise removes the simple exponential dependence of $f^d_k$ on depth, choosing $d_{\rm eff}$ correctly ensures that the learning procedure estimates an exponentially decaying Pauli fidelity that matches the data at, and only at, the correct depth $d$. This matching can be trivially achieved using a single depth learning sequence at depth $d$, but in practice due to SPAM and finite sampling issues we would require more than one depth point for learning. As mentioned in the main text, this close relationship between specifics of the learning procedure and mitigation performance is absent for Markovian noise. We suspect this stronger learning-mitigation dependence will be a feature of all non-Markovian noise distributions, though the exact nature of the relationship may not be as simple as it is in this case.

\subsection{Simulations of experimental quasi-static noise distributions}

\begin{figure*}[t!]
    \centering
    \includegraphics[width=1.7\columnwidth]{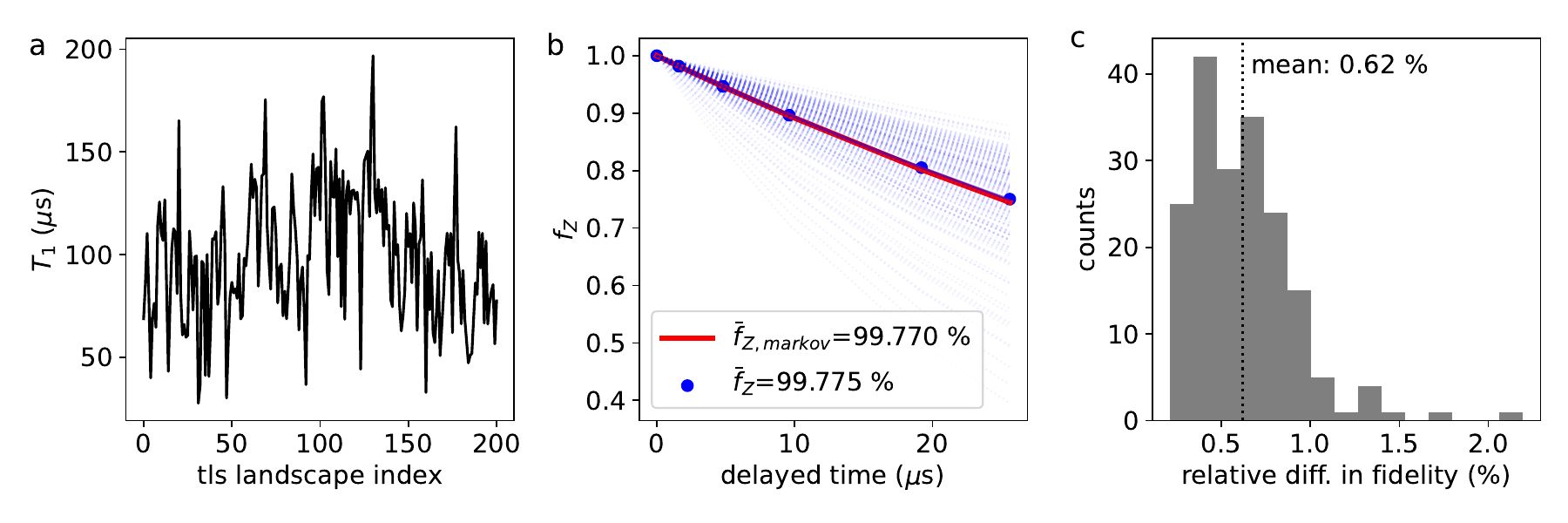}
    \caption{(a) An example $T_1$ distribution derived from an experimentally measured $\mathcal{P}_e$ distribution. (b) The estimated Pauli fidelity by adopting the $T_1$ distribution described in (a). We assume that each $T_1$ is drawn with equal probability to compute the ensemble average of the Pauli fidelity. The Pauli fidelity after $d$-fold repetition with a sampled $T_1$ value is indicated as a blue dotted line, and the ensemble average at each $d$ is indicated as a blue symbol. We extract the Pauli fidelity from an exponential fitting procedure to obtain an average Pauli fidelity, $\bar{f}_Z$. The Pauli fidelity computed with a Markovian assumption ($\bar{f}_{Z,Markov}$, red line) is shown for comparison. (c) The procedure in (b) is repeated for 32 different TLS landscapes for 6 qubits (192 TLS landscapes). Then we compute the relative difference in fidelity as $|\bar{f}_{Z,\text{Markov}}-\bar{f}_Z|/\bar{f}_{Z,\text{Markov}}$ for the 192 different landscapes, and the results are plotted as a histogram. }
    \label{fig:sims_fid}
\end{figure*}

Having explored the impact of quasi-static noise through simple models for the noise distributions, we now use the machinery developed in the previous subsections with $T_1$ distributions measured in our experiments. 

We first revisit Sec.~\ref{sec:learning} and examine the impact of quasi-static $T_1$ on Pauli fidelity estimation. We utilize Eq.~\ref{eq:repeat} and perform a simulation using experimentally measured $\mathcal{P}_e$ distributions. For simplicity, we assume that the only source of noise is $T_1$ decay. The  stochastic single-qubit Pauli fidelity for Pauli $Z$ is therefore $f_{Z,i}=e^{-t/T_{1,i}}$ where $T_{1,i}$ is $T_1$ measured for the $i$th qubit-TLS interaction landscape. We assume that the average noise channel is obtained from uniformly sampled $T_1$ over the measured $T_1$ values for varying $k_{TLS}$ at each shot. We obtain the $T_1$ distribution from the $\mathcal{P}_e$ distribution by ignoring SPAM error and using $P_1=e^{-t_1/T_1}$ where $t_1=40~\mu$s. 

Figure~\ref{fig:sims_fid}a shows an example $T_1$ distribution computed from experimentally measured $\mathcal{P}_e$ distribution. Finally, we estimate the average fidelity by using the assumed $T_1$ distribution in a simulation of the learning procedure described in Eq.~\ref{eq:repeat}. We set the unit time step $t_0=0.2\mu$s, and repeat the noisy identity operation with single-qubit Pauli channel for $d=0,8,24,48,96,128$. First, the average channel fidelity is obtained from $\bar{f}_{Z,\text{Markov}}=\sum_i (1/N) f_{Z,i}\simeq0.99770$ where $N$ is the number of sampled $T_1$ values. Under the full Markovian assumption, the corresponding fidelity for each repetition $d$ is simply computed from $\mathbb{E}[f_Z]^d=\bar{f}_{Z,\text{Markov}}^d$ as discussed in Eq.~\ref{eqn:circdecay}. The corresponding results are plotted as the red solid line in Fig.~\ref{fig:sims_fid}b. This is close to the assumed scenario during our fidelity extraction procedure, and is compared against the experimental scenario with quasi-static noise below.  

For the quasi-static case, the fidelity of each element of the distribution is computed using the given $T_{1,i}$ for each repetition. Each fidelity decay curve from the quasi-static distribution is depicted as a blue dotted line in Fig.~\ref{fig:sims_fid}b. Note that depth $d$ benchmark circuit will run with a given number of shots and the fidelity is estimated by averaging various quasi-static noise environments. Namely, the depth $d$ quasi-static average fidelity is obtained as $\mathbb{E}[f_Z^d]=\sum_i (1/N) f_{Z,i}^d\neq \bar{f}_{Z,\text{Markov}}^d$ and plotted as a blue circles in Fig.~\ref{fig:sims_fid}b. This is close to the experimental scenario. The average fidelity $\bar{f}_Z$ is estimated from an exponential curve fit to the data points indicated by the blue circles, giving $\bar{f}_Z\simeq0.99775$, which is not exactly the same but close to $\bar{f}_{Z,\text{Markov}}$. We repeat the above analysis for all our experimental data, consisting of 32 different qubit-TLS landscape for six qubits, or 192 different TLS landscapes. The relative difference in the Markovian average fidelity and the quasi-static estimate to the average fidelity is defined as $|\bar{f}_{Z,\text{Markov}}-\bar{f}_Z|/\bar{f}_{Z,\text{Markov}}$. Fig.~\ref{fig:sims_fid}c shows the distribution of this relative difference over 192 different TLS landscapes. The statistics show an average relative difference of $0.62\pm0.29\%$. This means that the fidelity extracted from the experimental protocol is reasonably close to the fidelity computed under the full Markovian assumption.

Now we revisit Sec.~\ref{app:issue-gen} and \ref{app:issue-mit} to
further investigate on the impact of our experimentally measured $T_1$ distributions on PEC error mitigation. As previously mentioned, for a single qubit with a fluctuating $T_1$ drawn from an unknown distribution, we can define a distribution of Pauli channels whose only source of error is $T_1$. Each member of this distribution has Pauli fidelities $f_{Z,i} = e^{-t/T_{1},i}$ and $f_{X,i} = f_{Y,i} = \sqrt{f_Z}$. For multiple qubits, if $T_1$ is the only error source, then the Pauli fidelity of higher weight Pauli fidelities will simply be the product of the corresponding weight-one Pauli fidelities. Focusing on the all Pauli $Z$ observable on $M$ qubits, the relevant Pauli fidelity is $\prod_n^M f_{Z_n,i}$.

To understand the impact of quasi-static $T_1$ it is sufficient to consider a circuit formed by repeated applications of an identity layer of duration $\tau$. For a noise instance $T^{(n)}_{1,i}$, where the superscript indicates the qubit, the expectation value of the all $Z$ observable at depth $d$ for this circuit is
\begin{align}
    \left<\bigotimes^M_n Z_n\right> = \prod_n f_{Z_n,i}^d = \exp\left(-\sum_n \frac{d\tau}{T_{1,i}^{(n)}}\right),
\end{align}
To simulate learning on 6 qubits using the above formula we calculate for each qubit the single-qubit expectation value $\left<Z_n\right>$ for all values of $T_1$ in the experimentally measured distribution. We then average over the distribution at each depth to obtain a single, generically non-exponential decay curve. We fit this decay to an exponential function to obtain a time constant we call $\bar{T}^{(n)}_1$ for each qubit. As in the experiments, we use a learning sequence of depths $\left[0, 4, 12, 24, 48, 64\right]$ and $\tau = 135$~ns.

\begin{figure}[t!]
    \centering
    \includegraphics[width=1\columnwidth]{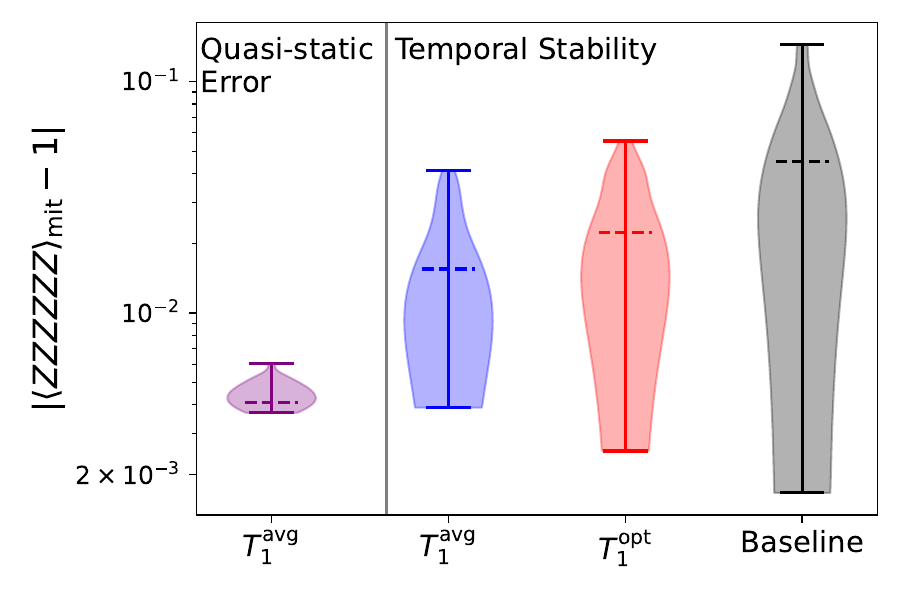}
    \caption{(Left panel) Distribution of quasi-static bias  over time slices for which we have experimentally measured $T_1$ as a function of $k_{TLS}$. Learning and mitigation are performed with the TLS distribution from the same time slice. (Right panel) Temporal stability of the different TLS modulation schemes, using learning data taken at $t=0$ to mitigate at all future times. The left violin-plot (blue) shows the performance of the average-channel protocol where $k_{TLS}$ is varied during both learning and mitigation; the middle (red) plot is for the optimized-channel protocol where the best $T_1$ point is used at each time. The right (black) plot shows the baseline experiment where a fixed $k_{TLS}$ value is used for both learning and mitigation.}
    \label{fig:sims}
\end{figure}

To simulate mitigation at a target depth of $24$, we calculate the expectation value $\left<ZZZZZZ\right>$ for all values of $T_1$ from each qubit's distribution, which amounts to fully sampling the multi-qubit product distribution. This gives a distribution of noisy expectation values, and we then mitigate each element of this distribution by dividing by the loss of signal expected from the results of our learning procedure, given by
\[
    \bar{f}_{ZZZZZZ} = \exp\left(-\sum^6_n \frac{24\tau}{\bar{T}_1^{(n)}}\right). \label{eqn:mitsim}
\]
Finally, we average over this mitigated distribution to obtain our expected mitigation performance under the modulation strategy where $k_{TLS}$ is varied both during learning and mitigation.

Since we have experimental data of the TLS landscape as a function of both $k_{TLS}$ and time (see Fig.~\ref{fig:t1}c) we can perform the above procedure at each time slice to study different experimental $T_1$ distributions. Using data from the same time slice for both learning and mitigation, we measure the bias error induced in mitigation by quasi-static $T_1$ by comparing the deviation of the mitigated $\left<ZZZZZZ\right>$ from its ideal value of one. A violin-plot of the distribution over time slices of this quasi-static induced bias error is shown in the left panel of Fig.~\ref{fig:sims}. As can be seen, simply having quasi-static $T_1$ induces a bias error with an average on the order of $4\times 10^{-3}$ (in absolute terms). We emphasize again that this error cannot be corrected by additional sampling, as evidenced by the fact that our simulation method is equivalent to the infinite sampling limit of PEC.

We can also study the temporal stability of the averaged learning procedure, by using the $\bar{T}^{(n)}_1$ calculated from the $t=0$ distribution to mitigate later time slices. The results of this are show in the left (blue) violin-plot of the right panel of Fig.~\ref{fig:sims}. As can be seen, both the average and spread of the bias error have increased considerably, indicating that \emph{for this data set} the $k_{TLS}$ variation of the TLS landscape is not a good proxy for its temporal variation. We also compare the averaged channel protocol to the optimized channel protocol, where $k_{TLS}$ is fixed to the value (at a given time) which gives the best $T_1$, again using $t=0$ learning to mitigate at later times. The temporal stability for this protocol is shown in the middle (red) violin-plot of Fig.~\ref{fig:sims}, which shows comparable, slightly increased, average and spread of the bias error compared to the averaged channel strategy. We also compare to the baseline experiment with no $k_{TLS}$ variation or optimization, where we take the $T_1$ value at the same, fixed value of $k_{TLS}$  at all times, and as before use the $t=0$ to mitigate later times. As can be seen, without variation or optimization the spread of performance and the average error both increase significantly.

\section{Discussions on experimental strategy} \label{app:expstrategy}
\subsection{Passive strategy without qubit-TLS landscape modulation}
\begin{figure*}[t!]
    \centering
    \includegraphics[width=1.7\columnwidth]{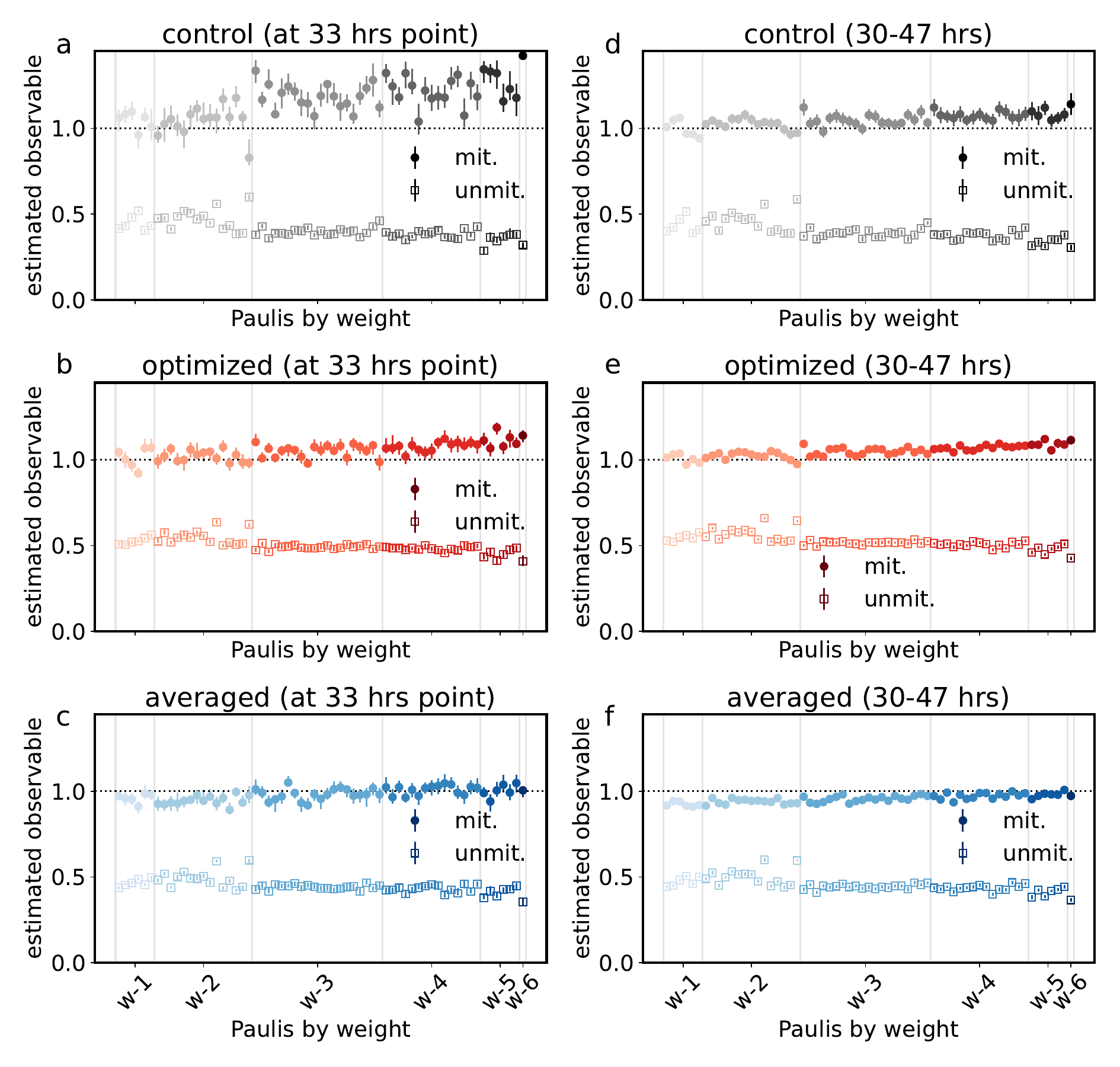}
    \caption{{\bf PEC results for all observables using one or multiple learn-mitigate events.} We use the same data points presented in Fig.~\ref{fig:mit}(a--c) and compute other possible $Z$ Pauli operators. We use 1 time point to show observables for various Pauli $Z$ operators for (a) control, (b) optimized, and (c) averaged noise channel. Similarly, we use averaged quantity over 10 time points for (d) control, (e) optimized, and (f) averaged noise channel. The plot shows the error mitigated (filled symbol) and unmitigated (open symbol) observable estimate as a function of observable weight. Error bars for the unmitigated and mitigated results in (a--c) are obtained by bootstrapping the PEC result 25 times. The standard error is plotted as an error bar for (d--f). The ideal value of 1 is indicated as a dotted horizontal line.}
    \label{fig:allweights}
\end{figure*}
The PEC procedure utilizes prior knowledge of the noise model from a noise learning experiment carried out before the error mitigation experiment. It is inevitable that there will be a time delay between the learning and mitigation events, and we thus expose our results to temporal fluctuation of the model over the delay time scale.

The observation of the model parameter fluctuation in Fig.~\ref{fig:model} reveals that there are catastrophic changes on top of relatively small temporal fluctuations. We also show that one of the largest changes in the model parameter is induced by strong qubit-TLS interactions as illustrated in Fig.~\ref{fig:model} b--d. The impact of such strong model parameter fluctuation is visible in  simple circuits, such as an unmitigated target circuit or the readout-only circuit (see e.g., Fig.~\ref{fig:controlonly}c, e). Defining appropriate monitoring circuits and interleaving them along with a target mitigation circuit provides a means to detect and exclude undesirable results~\cite{kim2023evidence}. However, we have a finite run-time budget for monitoring circuits, and thus the monitoring quantity is subject to have uncertainties due to limited resources, such as insufficient shot or twirling instances. Temporal fluctuations other than obvious catastrophic changes could therefore be challenging to detect using this scheme.

As a consequence, it may be beneficial to learn and mitigate for the target circuit multiple times, rather than learn and mitigate only once. 
For instance, Fig.~\ref{fig:mit}a highlights two periods of time where the mitigation values are fluctuating by a non-trivial amount, and Figure~\ref{fig:allweights}a shows an example of mitigated observables during that period of time. In a repeated learn-mitigate scheme, we would average the mitigation results over time, with the hope that this minimizes the impact of the strong temporal fluctuations. This is shown in Fig.~\ref{fig:allweights}d for many observables sorted by weight, e.g., weight-1 observables include $\braket{ZIIIII}$ and weight-5 observables include $\braket{ZZIZZZ}$. Points in Fig.~\ref{fig:allweights}d are obtained by averaging the repeated learn-mitigate experiments within one of the shaded regions of Fig.~\ref{fig:mit}a (ten points from $30-47$ hours). The averaged estimate from multiple learn-mitigate protocol provides a reasonable mitigated observable estimate even though the outcomes of the individual mitigation experiments may be unreliable.

The repeated learn-mitigate strategy may work under the assumption that the underlying distribution of possible noise channel configurations is temporally invariant, such that at each time we sample a specific noise configuration for learning, and sample a (potentially) different configuration for mitigation. However, such a case still can introduce a bias to our mitigated estimate even in the infinite noise configuration sampling limit. To understand why this is the case, consider the simple case of mitigating a Clifford circuit, which in the infinite shot limit can be represented by $\left<O\right>_{mit} = \left<\tilde{O}\right>/f_{pred} = f_{exp}\left<O\right>/f_{pred}$. As before, $f_{pred}$ and the $f_{exp}$ are the circuit fidelity computed from the noise channel configuration at learning and during the mitigation experiment respectively. The repeated learn-mitigate strategy estimates an expected value of $\left<O\right>_{mit}$, given by 
\begin{align}
    \mathbb{E}\left[\left<O\right>_{mit} \right] = \mathbb{E}\left[\frac{f_{exp}}{f_{pred}}\right] \left<O\right>.
\end{align}
Crucially, even though $f_{pred}$ and the $f_{exp}$ are drawn from the same distribution, the expected value of their ratio is not equal to the ratio of their expected values, i.e.~$\mathbb{E}\left[f_{exp}/f_{pred}\right] \neq \bar{f}_{exp}/\bar{f}_{pred}$, but is a function of higher-order moments of the distribution of noise configurations. This introduces the aforementioned bias to the result of the repeated learn-mitigate strategy.

Nevertheless, we believe this strategy provides a more stable estimate compared against a mitigation attempt based on a single pair of noise configurations for learning and mitigation. Of course, the additional experiment time for the repeated learn-mitigation stregegy comes at the expense of stability. While we have a certain degree of control over the TLS landscape for optimized or averaged noise channel, it is not yet possible to have a complete control of the TLS landscape. Therefore, the repeated learn-mitigate strategy would benefit all three strategies, yet the required number of repetitions would be smaller for optimized and averaged noise channel than that of the control noise channel. 

\subsection{Importance of qubit-TLS landscape modulation strategy}

\begin{figure*}[!t]
    \includegraphics[width=1.4\columnwidth]{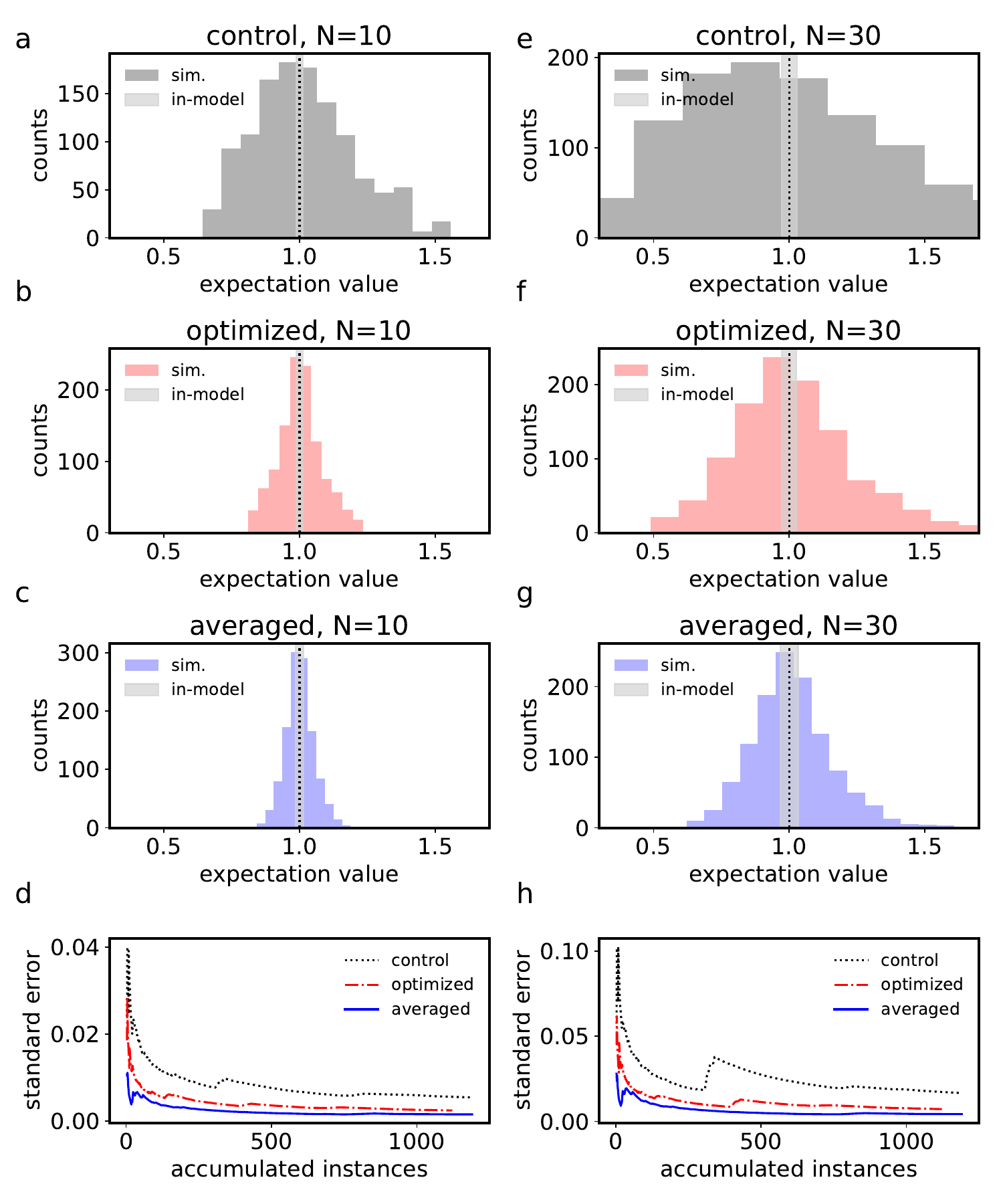}
    \caption{{\bf Predicted distribution for mitigated observable at larger depth.} We use the learned noise model to predict the mitigated observable distribution. The prediction is carried out by estimating the fidelity based on $n_1$th and $n_2$th learned noise models ($n_1\neq n_2$). The fluctuation between two separate learned noise models will be reflected as a deviation of the predicted observable, $\braket{O}_{mit,n}^{pred}$, from its ideal value of 1 (see detailed definition in the text). The predicted mitigated observable spread is plotted as filled bars for (a) control, (b) optimized, and (c) averaged noise channel experiments. Here, we estimate weight-6 observable, $\braket{ZZZZZZ}$, of the benchmark circuit in Fig.~\ref{fig:circuit} with depth $N=10$. This is the same circuit condition as the error mitigation experiments described in Fig.~\ref{fig:mit}. (d) As we proceed with the repeated learn-mitigation strategy, we accumulate results from $m$ separate learning instances (x-axis) to better estimate the observable. Here, we plot the standard error of the estimated observable, $\bar{\sigma}^{pred}_{mit}$ (y-axis), which is defined in Sec.~\ref{app:expstrategy}. The control and optimized cases show that the uncertainty does not decrease gradually when we encounter relatively large fluctuations at certain learning instances; instead it shows a sudden jump. An even more exaggerated trend is observed for a deeper circuit with $N=30$ in (e--h). (a--c) and (e--g) show the standard deviation of the model uncertainty as grey bars. The model uncertainty is obtained by bootstrapping the learning experiments for 100 times for each learning event. For a given $n$th learned noise model experiment, we carry out $\braket{O}_{mit,n,b}^{model}=f_{pred,n,b+1}/f_{pred,n,b}$ where $b\in\{1,3,5,\cdots,99\}$ is bootstrap index. }
    \label{fig:histdeeper}
\end{figure*}

Although the repeated learn-mitigation strategy may be effective for short-depth circuits, we observe that large fluctuations eventually limit the level of uncertainty we could achieve for deeper (or larger) circuits. To take a closer look at this issue, we utilize the obtained time series of the learned noise model to predict fluctuations on the mitigated observable. We compute the circuit fidelity for the benchmark circuit in Fig.~\ref{fig:circuit}, $f_{pred,n_1}$, for $n_1$th learned noise model where $n_1\in\{1,2,\cdots,N_{l}\}$ for $N_l$ total number of learning experiment attempts. Then, we define mitigated observable as $\braket{O}_{mit,n}^{pred}=\braket{O} f_{pred,n_1}/f_{pred,n_2}=f_{pred,n_1}/f_{pred,n_2}$, where an ideal observable is $\braket{O}=1$ for the benchmark circuit and $n\in\{1,2,\cdots,N_l(N_l-1)\}$ is an index to represent $(n_1,n_2)$ that satisfies $n_1\neq n_2$. This consideration assumes that our model is uncorrelated in time and representative, and we are merely probing fluctuations between two separate learning events. 
To provide a more concrete picture, we plot model uncertainty as a filled gray region, which reflects the one standard deviation uncertainty associated with model learning protocol described in Sec.~\ref{app:controlonly}. All experiments show a distribution whose spread is larger than the model uncertainty. This confirms that the observed variation in model parameters cannot be explained by uncertainty of the noise learning protocol, but likely come from \emph{actual} fluctuations in the noise. In this case, learning and mitigating once may end up giving us a mitigated observable estimation deviated larger than than learning protocol uncertainty due to the temporal fluctuation of the model. As described in the previous section, we may improve this by accumulating multiple learning-mitigation events. The confidence of our estimate in this case is described by Figure~\ref{fig:histdeeper}d as as one standard error deviation of the averaged observable prediction.  
Namely, the averaged observable prediction is obtained by accumulating $m$ repeated learning routines as $\braket{\bar{O}}_{mit}^{pred}=(1/m)\sum_{n=1}^{m}\braket{O}_{mit,n}^{pred}$ where $m$ could be effectively considered as the number of repeated learn-mitigate experiments. The standard error is then computed as $\bar{\sigma}^{pred}/\sqrt{m}$ where
$\bar{\sigma}^{pred}=\left[ (1/m) \sum_{n=1}^m (\braket{O}_{mit,n}^{pred}-\braket{\bar{O}}_{mit}^{pred})^2 \right]^{1/2}$. The standard error is expect to monotonically decrease, and the averaged noise channel in Fig.~\ref{fig:histdeeper}d shows such a monotonic decrease. However, the control and optimized noise channels show jumps that significantly elevate the standard error. These discontinuities correlate with relatively large noise fluctuations illustrated in Figs.~\ref{fig:model}b and ~\ref{fig:model}c for the control and optimized noise channels, respectively. The trend is more obvious when we proceed towards deeper circuits for error mitigation. Figure~\ref{fig:histdeeper}e--g illustrates distribution of the predicted mitigated observable but for a circuit that is three times deeper than the benchmark circuit considered in Fig.~\ref{fig:histdeeper}a--c. Note that the spread of the mitigated observable is significant for the control experiment. As a result, the large fluctuations in the noise model results in larger jumps in the standard error estimation in Fig.~\ref{fig:histdeeper}h. Since we have a limited budget of repetitions of the learn-mitigation protocol, this is critical for deeper circuits where the sampling cost is exponentially higher.
In the regime of problems that are intractable with classical methods, the circuits tend to be deep and cover a large number of qubits. Here, it is especially important to have a stable noise model that could results in a monotonic decrease in standard error so that one could obtain mitigated observable with desired uncertainty using a reasonable amount of resources.

\section{Additional discussions for the learning experiments} \label{app:controlonly}

\begin{figure}[!ht]
    \includegraphics[width=1.0\columnwidth]{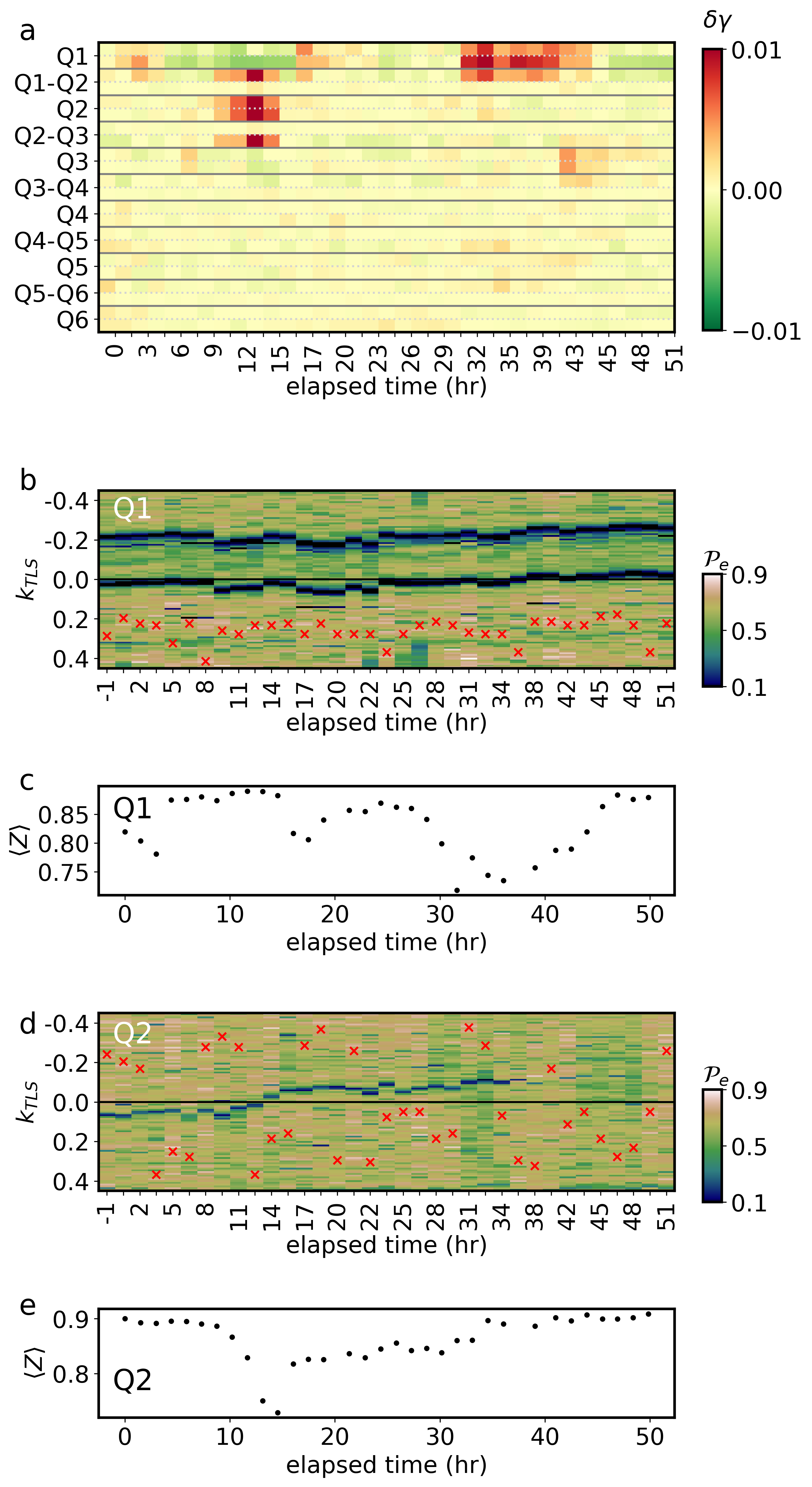}
    \caption{{\bf Temporal noise model fluctuations for the control experiment.} (a) We plot the qubit- and pair-wise sampling overhead for the control experiment again for clarity. The blocks along the y-axis represent the different qubits and qubit-pairs present in the model. Each block contains two rows, representing deviations in the aggregated local sampling overheads from their median value over the duration of the experiment, for layers 1 and 2, respectively. (b) The qubit-TLS interaction landscape for Q1. (c) Readout mitigation circuit interleaved with the mitigation circuit. $\braket{Z}$ is estimated for Q1. (d) Qubit-TLS interaction landscape for Q2. (e) Readout mitigation circuit interleaved with the mitigation circuit. $\braket{Z}$ is estimated for Q2. }
    \label{fig:controlonly}
\end{figure}

Here, we intentionally focus on the control experiment and describe how a large fluctuation event manifests itself in a simple monitoring circuit.

Fig.~\ref{fig:controlonly}a provides a detailed picture of model coefficient fluctuation by computing a \emph{one-local} sampling overhead per qubit, $\gamma_{l,q}=\exp\left(2\sum_{k\in\mathcal{K}_q}\lambda_k\right)$, where $q$ and $l$ are qubit and layer indices, respectively, and $\mathcal{K}_q$ is the set of weight-1 Pauli operators associated with qubit $q$. Likewise, a sampling overhead per pair can be defined as $\gamma_{l,q_1q_2}=\exp\left(2\sum_{k\in\mathcal{K}_{q_1q_2}}\lambda_k\right)$, where $\mathcal{K}_{q_1q_2}$ is the set of weight-2 Pauli operators associated with a pair of qubits $q_1$ and $q_2$. We use the local sampling overhead parameters as a coarse scalar metric to track changes in the noise model parameters.

Figure~\ref{fig:controlonly}a illustrates strong qubit-TLS interactions that randomly occur for Q1 and Q2. While the optimized noise channel may be able to avoid such interactions for the next round of optimization, the control experiment relies on the scenario that the strong interaction diffuses away from the qubit frequency. As a result, the control noise channel may sometimes be exposed to a strong qubit-TLS interaction for an extended period of time, as highlighted by red regions in Fig.~\ref{fig:controlonly}a. The observation is consistent with the strong qubit-TLS interaction signaled by a dip in $\mathcal{P}_e$ (dark color) at neutral point, $k_{TLS}=0$. Figure.~\ref{fig:controlonly}b shows that Q1 has persisting qubit-TLS interaction for the entire monitoring period. As a result, both the qubit- and pair-wise model parameters associated with Q1 fluctuate more than other parameters. Figure~\ref{fig:controlonly}d describes a clear instance where a strong qubit-TLS interaction drifts in $k_{TLS}$ parameter space and hits the neutral point near the $\sim13$ hours time point. This also shows a good correlation with the qubit- and pair-wise model parameter fluctuations associated with Q2 around $\sim 13$ hours.

While the above example probes strong qubit-TLS interaction during the learning procedure, we could also probe such events during the mitigation procedure. An unmitigated observable from the target circuit is one candidate as it has a natural relationship with the mitigated observable. However, the benchmark circuit at this depth entangles every qubit and it is therefore challenging to determine which qubits may be problematic based on individual observables. Instead, we interleave a circuit that does not have entanglement so that strong qubit-TLS interaction can be identified by a particular local observable.
Readout-error mitigation circuits~\cite{berg2022model} can be used as an easy-to-interpret monitoring probe, that requires no additional run-time: we prepare $\ket{0}$ state for each qubit and randomly apply an \gates{X} or \gates{I} gate followed by a projective measurement in the Z-basis. Although intended for readout error mitigation, we also can use the circuit to monitor for particularly strong qubit-TLS interactions. In the absence of entangling gates, the local observable $\braket{Z}$ of each qubit may serve as a metric that monitors a quality of the single-qubit gates. When a strong qubit-TLS interaction is present and the RF control signal for an \gates{X} gate is not properly delivered to a qubit, the gate may partially flip the qubit, resulting in a deviation of $\braket{Z}$ estimation from its ideal value of 1. Figure~\ref{fig:controlonly}c and~\ref{fig:controlonly}e shows $\braket{Z}$ observable of the readout only circuit for Q1 and Q2, respectively. We observe a particularly strong qubit-TLS interaction that arises and affects Q1 and Q2 (at different times), signaled by a dip in $\braket{Z}$. The large dip correlates with the large fluctuation observed in Fig.~\ref{fig:controlonly}a.


\end{document}